\documentclass[floatfix,showpacs,aps,amsmath,amssymb,nofootinbib,twocolumn]{revtex4}

\usepackage{graphicx}
\usepackage{color} \usepackage[tight]{subfigure}

\newcommand{\abs}[1]{\lvert#1\rvert}

\newcommand{\E}{\mathcal{E}}
\renewcommand{\L}{\mathcal{L}}
\newcommand{\W}{\mathcal{W}} 
\newcommand{\ang}[1]{\overset{\circ}{#1}}
\newcommand{\Eo}{\ang{\E}} \newcommand{\Lo}{\ang{\L}}

\newcommand{\FT}[1]{\mathcal{F}_{T}[#1]}
\newcommand{\FTinv}[1]{\mathcal{F}^{-1}_{T}[#1]}

\newcommand{\Lb}{\mathbf{\L}}
 \newcommand{\Lob}{\ang{\Lb}}

\newcommand{\alphab}{\mathbf{\alpha}}

\newcommand{\tri}{\triangle}

\newcommand{\tif}{\widetilde{f}}

\newcommand{\Tintx}{\frac{1}{T}\int_0^T dx\,}
\newcommand{\TTinv}{\frac{1}{2T}}
\newcommand{\Tinv}{\frac{1}{T}}
\newcommand{\iinv}{\frac{1}{i}}
\newcommand{\half}{\frac{1}{2}}
\newcommand{\Ph}{\widehat{P}}

\newcommand{\suminfn}{\sum_{n=0}^\infty}
\newcommand{\suminfm}{\sum_{m=0}^\infty}

\newcommand{\infact}{\frac{i^n}{n!}}
\newcommand{\imfact}{\frac{i^m}{m!}}

\newcommand{\gamfrac}{\frac{\gamp_1}{\game_1}}

\newcommand{\m}[1]{\langle #1 \rangle}

\newcommand{\del}[2]{\frac{\partial #1}{\partial #2}}

\newcommand{\gamp}{\gamma^\phi}
\newcommand{\game}{\gamma^E}
\newcommand{\gamej}{\gamma^{E_j}}

\newcommand{\var}{\text{cov\,}}

\newcommand{\tforall}{\quad \text{for all }}

\newcommand{\figwidth}{2.8in}

\begin{document}
\title{Random Walks for Spike-Timing Dependent Plasticity} 
\author{Alan Williams}
\email{williaal@ohsu.edu}
\affiliation{Neurological Sciences
  Institute, Oregon Health \& Science University, 505 NW 185th 
Avenue, Beaverton, OR 97006}
\author{Todd K. Leen} 
\email{tleen@cse.ogi.edu}
\affiliation{Department of
  Computer Science and Engineering, OGI School of Science \&
  Engineering, Oregon Health \& Science University}
\author{Patrick D. Roberts}
\email{robertpa@ohsu.edu}
 \affiliation{Neurological Sciences
  Institute, Oregon Health \& Science University, 505 NW 185th 
Avenue, Beaverton, OR 97006 }

\pacs{87.18.Sn,87.19.La,75.10.Nr} 

\date{\today}

\begin{abstract}
Random walk methods are used to calculate the moments of negative image 
equilibrium distributions in synaptic weight dynamics governed by 
spike-timing dependent plasticity (STDP). The neural architecture of 
the model is based on the
electrosensory lateral line lobe (ELL) of mormyrid electric fish,
which forms a negative image of the reafferent signal from the fish's
own electric discharge to optimize detection of sensory electric
fields. Of particular behavioral importance to the fish is the 
variance of the equilibrium postsynaptic potential in the presence of 
noise, which is determined by the variance of the equilibrium weight 
distribution. Recurrence relations are derived for the moments of the 
equilibrium weight distribution, for arbitrary postsynaptic potential 
functions and arbitrary learning
rules. For the case of homogeneous network parameters, explicit closed form solutions are developed for the covariances of the synaptic weight and postsynaptic potential distributions.

\end{abstract}

\maketitle

\section{Introduction}\label{sec.intro}

Activity dependent synaptic plasticity is believed to be a fundamental mechanism for
learning and adaptation in neural systems. 
\cite{Hebb49}. Experimental observation of plasticity depending on mean spike rate \cite{Lomo71,Bliss73} led to rate-based models, in which the changes in synaptic weight depend on correlations in the mean spike rate of presynaptic and postsynaptic cells \cite{Sejnowski77,Bienenstock82}. Since mean spike rates are necessarily averages
over time windows containing many spikes, the timing of individual
spikes is ignored in rate-based models. More recent experimental
work \cite{Markram97a,Bell97a,Bi98} has shown that in some systems
plasticity does depend on the precise timing of individual spikes. Models of such \emph{spike-timing dependent
  plasticity} (STDP) \cite{Abbott00} assume the weight change due to each presynaptic and postsynaptic spike pair is given by some function of the time between them, called the spike-timing dependent \emph{learning rule} \cite{Gerstner96,vanRossum00,Rubin01,Yoshioka02,Zhigulin03,Cateau03}. Changes due to all pairs of presynaptic and postsynaptic spike pairs are then summed to give the weight change due to presynaptic and postsynaptic spike trains. 
  
One system in which STDP has been found experimentally is the electrosensory lateral line lobe (ELL), a cerebellum-like structure in mormyrid electric fish \cite{Bell97a}. The mormyrid
detects objects in its environment by emitting a pulsed
electrical discharge and observing the perturbations to the resulting
electrical field at the skin surface due to sensory objects. To
null out the predictable sensory input due solely to its own discharge, the
mormyrid employs an efference copy of the motor command which initiates the discharge. An array of time-delayed, time-locked
copies of the motor command  innervates medium ganglion
(MG) cells in ELL through plastic synapses. The MG cells also receive primary
afferent input from electroreceptors on the skin, through nonplastic synapses. The plastic synapses whose input is time-locked to the motor command enable the formation and maintenance of a negative image
\cite{Bell97b} of the primary afferent signal, via a spike-timing
dependent learning rule. The negative image effectively nulls out, in the MG cells, the
sensory effect of the fish's own discharge, which simplifies the detection of perturbations due to sensory objects. Plasticity
is critical to maintaining the negative image during ongoing changes in the
precise form of the discharge due to large daily or seasonal fluctuations in water
conductivity, or to changes in body size and shape during growth and development.

In order for the negative image to be maintainable in this way, the synaptic weight configuration giving rise to the negative image must be a stable 
equilibrium for the
mean weight dynamics induced by the spike-timing dependent 
learning
rule. Conditions for existence and stability of such negative image equilibria were
first explored in \cite{Roberts00b}, and extended to arbitrary spike-timing dependent 
learning rules and arbitrary 
postsynaptic potential functions in \cite{williams03}. 

The equilibrium weight \emph{distribution} in the presence of noise 
is also behaviorally important. Fluctuations in the weights due to 
noise lead to fluctuations in the negative image. For example, we will show in this paper that the variance of the equilibrium weight distribution is 
proportional to learning rate (i.e. to the magnitude of the weight 
changes induced by individual spikes or spike pairs).  A slow 
learning rate leads to a small variance in equilibrium weight 
distribution and hence a more accurate negative image; a fast 
learning rate gives a large variance in equilibrium weight 
distribution and a less accurate negative image. Detectability of 
sensory objects is improved by a more accurate negative image; thus 
to optimize detectability the learning rate should be slow. However, 
if the fish's own discharge is changing (due to changes in water 
conductivity or body shape, for example) then the negative image must 
be updated to remain accurate. Such adaptability of the negative 
image favors a fast learning rate, to allow the negative image to 
keep up with changes in the discharge. The twin requirements of 
detectability and adaptability are thus in direct conflict: any one 
choice of learning rate represents a compromise between them. A natural hypothesis is that the learning rate in mormyrid ELL is the slowest 
learning rate sufficient to provide adaptability of the negative 
image on timescales over which the fish's discharge varies in the 
wild. A faster rate would not significantly improve adaptability and 
would degrade detectability; a slower rate would unacceptably degrade 
adaptability.  

In the present paper we seek to lay the groundwork for the analysis of such issues in a rigorous mathematical fremework, by deriving analytic expressions for the moments 
of the equilibrium weight distribution (when it exists) for arbitrary spike-timing 
dependent learning rules and arbitrary postsynaptic potential 
functions. We work with a model based on mormyrid ELL, but the 
technique is applicable to any network architecture. The approach 
used is to express the weight dynamics as a discrete time, 
inhomogeneous random walk. From the master equation of this walk we 
derive a differential equation for the Fourier transform of 
the equilibrium weight distribution. Taylor expansion of this 
equation yields recurrence relations for the moments. 

Random walks have been used extensively to model other physical 
systems (see the bibliography \cite{liyanage}), and a large body of mathematical technique has been developed for their analysis \cite{hughes}. But they have not previously been applied to STDP, where the standard 
approach has been to use the Fokker-Planck equation \cite{vanRossum00,Rubin01,Cateau03}. Given that the Fokker-Planck equation is at best an approximation\footnote{Moreover, the conditions under which the approximation is a good one, especially for the nonlinear Fokker-Planck equation, are far from clear \cite{vankampen}. Further discussion of this issue, in the context of STDP, will be the subject of a future paper.} when applied to discrete stochastic processes \cite{vankampen}, whereas random walk methods are exact, we believe it would be prudent to explore the utility of random walk methods for the analysis of STDP.

The structure of the paper is as follows. In Section \ref{sec.walks} we summarize the background facts about random walks, master equations, and characteristic functions that will be used in the present paper. In Section \ref{sec.framework} we describe the architecture and dynamical assumptions of the model, and in Section \ref{sec.dynamics} we derive the random walk for the weight dynamics, for arbitrary system parameters. In Section \ref{sec.one} we illustrate the method for deriving recurrence relations for the moments of the equilibrium weight distribution by applying the method in the simplest possible setting, the case of a single synaptic weight. We then in Section \ref{sec.mult} apply the method to the full architecture, with arbitrary system parameters. In Section \ref{sec.hom} we specialize to the case of homogeneous system parameters, deriving more explicit analytical results for the covariance of the equilibrium weight and postsynaptic potential distributions. Finally in Section \ref{sec.examples} we compute the weight and postsynaptic potential covariances for several examples of biological interest, and compare our predictions with Monte Carlo simulations.

\section{Random Walks, Master Equations, and Characteristic Functions}\label{sec.walks}

The term \emph{random walk} refers to any stochastic process 
in which the state variables change only at discrete times. 
The changes in state variables are called \emph{steps}; from any given position there is a set of possible steps, each having a certain probability (or probability density). The set of possible steps may be discrete or continuous, and both the step values and step probabilities may depend on position. 

Random walks are natural models for systems having temporally 
discrete dynamics. They are natural models for STDP because weight changes in STDP are due to temporally discrete events (spikes or spike pairs).

Suppose a state variable $w$ undergoes a random walk. Let the possible steps from position $w$ be $j_{w}(x)$, for $x$ in some index set $X$. Let the step $j_{w}(x)$ occur with probability density $\rho_{w}(x)$ in $x$. Let $P_{n}(w)$ be the probability distribution for $w$ after $n$ steps. We wish to derive the equation of motion for $P_{n}(w)$, usually referred to as the \emph{master equation}. 

If the state variable is $w'$ after $n$ steps and $w$ after $n+1$ steps, then $w=w'+j(x,w')$ for some $x$. The probability for the state variable to be between $w$ and $w+dw$ after $n+1$ steps is therefore
\begin{equation}
\label{ }
P_{n+1}(w) dw = \int dx \, \rho_{w}(x) \bigl[ P_{n}(w') dw'\bigr]. \notag
\end{equation}
Hence the master equation is
\begin{equation}
\label{}
P_{n+1}(w) = \int dx \, \rho_{w'}(x) P_{n}(w') \frac{dw'}{dw} . \notag
\end{equation}
The quantity $dw'/dw$ compensates for any change in the density of states from time $n$ to time $n+1$, due to position dependence of the set of step values. From $w=w'+j(x,w')$ we have
\begin{equation}
\label{dwdw'}
\frac{dw'}{dw} = \frac{1}{1+\frac{\partial}{\partial w'} j_{w'}(x)},
\end{equation}
and hence the master equation is
\begin{equation}
\label{mastergeneral}
P_{n+1}(w) = \int dx \, \rho_{w'}(x) P_{n}(w') \frac{1}{1+\frac{\partial}{\partial w'} j_{w'}(x)}.
\end{equation}
Suppose the set of step values is independent of position; then $\partial j_{w'}(x)/\partial w'=0$, and the density of states factor in the master equation is identically $1$. Denoting by $j(x)$ the common set of step values, we also have $w'$ explicitly in terms of $w$ and $x$: $w'=w-j(x)$. For such walks the master equation takes the simpler form
\begin{equation}
\label{masterspecial}
P_{n+1}(w) = \int dx \, \rho_{w-j(x)}(x) P_{n}(w-j(x)) .
\end{equation}
All walks considered in the present paper will turn out to be of this type. 

A probability distribution $P(w)$ is an equilibrium (stationary) distribution for the random walk if $P_{n}=P$ implies $P_{n+1}=P$; in other words, the dynamics of the walk leave $P$ unchanged. Hence $P(w)$ is an equilibrium distribution for the walk Eq. \eqref{masterspecial} if and only if it satisfies
\begin{equation}
\label{eqbmspecial}
P(w) = \int dx \, \rho_{w-j(x)}(x) P(w-j(x)) .
\end{equation}
To calculate the moments of a probability distribution $P(w)$, we will find it useful to invoke a property of its Fourier transform (often referred to as the \emph{characteristic function})
\begin{equation}
\label{charfngeneral}
\Ph(k)=\int dw \, e^{ikw} P(w).
\end{equation}
Taking the derivative with respect to $k$ in Eq. \eqref{charfngeneral} and evaluating at $k=0$ yields
\begin{eqnarray}
\label{derivsofchar}
\frac{d^{n}\Ph(k)}{dk^{n}} \bigr|_{k=0}& = & \Bigl( \int dw \, (iw)^{n} e^{ikw} P(w) \Bigr)  \bigr|_{k=0}  \notag \\
 & = & i^{n} \int dw \, w^{n} P(w)  \notag \\
 & = & i^{n} \m{w^{n}}.
\end{eqnarray}
Hence the moments of $P(w)$ are, up to powers of $i$, just the derivatives of the characteristic function $\Ph(k)$ evaluated at $k=0$. 

For further background on random walks, see \cite{hughes}.

\section{Framework}\label{sec.framework}

\begin{figure}
\begin{center}
\includegraphics[width=1.5in]{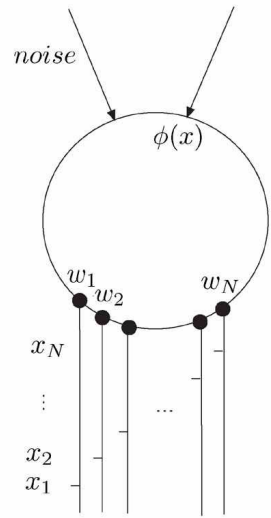}
\caption{Schematic of the architecture. The postsynaptic cell receives
  inputs from $N$ presynaptic neurons, a repeated sensory input
  $\phi(x)$, and a noisy input. Presynaptic cell $i$
  spikes at time $x_i$ in each period of $\phi$, and has synaptic
  weight $w_i$ onto the postsynaptic cell. }
\label{figschematic}
\end{center}
\end{figure}

The model consists of a single postsynaptic cell (representing an MG
cell) driven by a repeated sensory input (primary sensory reafference), an array of presynaptic cells whose spikes are time-locked to the repeated sensory input (the efference copy of the motor command), and noise (representing other unspecified inputs) \cite{Kempter99a,
  Roberts99a, Roberts00a} (Fig. \ref{figschematic}). This basic architecture is derived from mormyrid ELL, but is sufficiently general to capture the
dynamics of other neural systems hypothesized to have an array of
time-delayed, time-locked inputs through plastic synapses \cite{Hahnloser02,Ehrlich97}.

The framework for the neural dynamics is the spike response (SR) model
\cite{Gerstner93,Gerstner95}, without refractoriness. In SR models the effect of a presynaptic spike on a postsynaptic cell is add to the postsynaptic membrane potential a contribution given by the product of the synaptic weight and a postsynaptic potential function (PSP), which is a function of time after the spike. Spike response models include leaky integrate-and-fire (LIF) models as a special case \cite{Gerstner95}, and are used here because they simplify the derivation of analytic results.

The repeated sensory input is the postsynaptic potential (PSP) in the postsynaptic cell due to primary sensory afferents, over a single EOD sweep. Each time-locked presynaptic cell $i$ spikes (exactly once) at a fixed time within
each sweep of the repeated sensory input, causing a corrsponding PSP in the postsynaptic cell.

The total membrane potential in the postsynaptic cell is the sum of
the repeated sensory input, the noisy input, and the PSPs due to time-locked presynaptic spikes, weighted by synaptic efficacies (weights) $w_i$. This membrane potential causes the postsynaptic
cell to generate broad dendritic spikes\footnote{The postsynaptic cell also generates simple spikes, but these are not relevant for plasticity and no use is made of them in the present model. In this paper the phrase "postsynaptic spike"  refers solely to broad, dendritic spikes.} at a certain (noisy) rate. We assume that each presynaptic spike causes
a constant change in the weight $w_i$ (nonassociative learning), and each
postsynaptic and presynaptic spike pair causes a change in $w_i$
according to a spike-timing dependent learning rule, i.e. a function
of the time difference between the postsynaptic and presynaptic spikes
(associative learning).

The repeated sensory input has the form of a stereotyped pulse  
with variable interpulse interval. It has been found that the time-locked inputs occur for
approximately the duration of the pulse, and are absent during
interpulse intervals \cite{Bell97a}. The events which affect
plasticity are thus restricted to the duration of the pulses,
provided the width of the learning rule is much less than the width of
a pulse (a requirement we will impose below). When calculating the weight changes due to plasticity we may therefore omit
the variable interpulse intervals, and replace the repeated sensory input by a
\emph{periodic} input obtained by concatenation of the pulses.

Let the resulting period (pulse width) be $T$, and introduce two
time variables: $x\in [0,T)$ for the time within each period of
  the sensory input, and $t=nT$, $n\in \mathbb{Z}$ for the time of
  initiation of each such period \cite{Roberts99a, Roberts00a,
    Roberts00c}. General dynamical quantities will be functions of the
  pair $(x,t)$. The time-locked 
  presynaptic cell $i$ spikes at a fixed time in each period. Denote this time by $x_{i}$. Let $w_i(x,t)$ be the synaptic weight of presynaptic cell $i$, and let $\E_{i}(s)$ be the PSP evoked by
  a spike in cell $i$ at time $s$ after the spike. We will assume $\E_{i}$ is causal:
  $\E_{i}(s)=0$ for $s<0$. Let $\alpha_{i}$ be the nonassociative weight
  change due to a presynaptic spike by cell $i$, and $\L_{i}(s)$ the associative
  weight change due to a postsynaptic spike at time $s$ after a
  presynaptic spike by cell $i$. Let $\phi(x)$ be the periodic sensory input, and
  $U(x,t)$ the total postsynaptic potential due to the non-noisy
  inputs.

  We will assume that in each period of $\phi$, either zero or one postsynaptic 
spike occurs. The probability density (in $x$, for a given $t$) for a postsynaptic spike to occur at 
$(x,t)$ is assumed to be $\Tinv f(U(x,t))$, for
  some positive and strictly increasing function $f:\mathbb{R}\to 
[0,1]$. The probability of zero postsynaptic spikes in the period 
beginning at $t$ is then $1-\Tintx f(U(x,t))$. 
  Heuristically, the function $f$ is the effective
  gain function of the postsynaptic cell in the presence of the noisy
  inputs, with the maximum slope of $f$ indicating the noise level: high or low noise correspond to an $f$ with small or large
  maximum slope respectively. 
  
  We assume that the period of $\phi$ is sufficiently long that refractoriness can be ignored. In each period there is exactly one spike by each presynaptic cell and at most one spike by the postsynaptic cell, so if the period of $\phi$ is longer than the refractory period of all cells involved then refractoriness is irrelevant and can be omitted from the model.

We will implement changes in weights as discrete steps with no
internal time course. We update weights synchronously, once per sweep of the periodic sensory input, at time $x=0$ for each $t=nT, n \in
\mathbb{Z}$. The value of $w_i$ in the period beginning at $(0,t)$ is
then independent of $x$, and will be denoted $w_i(t)$. For synchronous
updating to be a good approximation, weight
changes per cycle must be small relative to the weights themselves -- the \emph{slow learning rate} assumption. Changes in weights due to different spikes and spike
pairs will be summed linearly.

In biological systems, synaptic weights have bounded magnitude and never change sign (Dale's Law). We impose no such boundary conditions in the
present model, but the results still apply to the biological case provided the
weight equilibria and equilibrium variances are such that weights are almost always in the region enclosed by biological bounds.

To simplify the derivation of the weight dynamics, we will assume that $\E_{i}(s),\L_{i}(s)$ are zero or negligible for
$\abs{s}>\tau_E,\tau_L$ respectively, with $\tau_E,\tau_L \ll T$. We
will also impose the slow learning rate assumption: $T \ll \tau_w$, where
$\tau_w$ is the time-scale over which weights undergo significant
relative change. The existence of approximate negative image
states requires \cite{williams03} that the spacing of presynaptic spike times be much
smaller than the widths of $\E_{i}$ and $\L_{i}$: $\delta \ll
\tau_E,\tau_L$. These time-scale assumptions can be summarized as
\begin{equation}
\label{time-scales}
\delta \ll (\tau_E,\tau_L) \ll T \ll \tau_w.  \notag
\end{equation}   
Typical values from mormyrid ELL are: $\delta < 1\text{ms}$
\cite{Bell92a}, [C.C. Bell, private communication], $\tau_E \sim
20\text{ms}$ \cite{Bell97a}, $\tau_L \sim 40\text{ms}$ \cite{Bell97a},
$T \sim 80\text{ms}$ [C.C. Bell, private communication], $\tau_w \sim
10^2T$ \cite{Bell97a}.

\section{Weight Dynamics}\label{sec.dynamics}
We now derive the random walk for the weight dynamics, by computing 
the possible weight changes $\tri w_i(t)=w_i(t+T)-w_i(t)$ and their 
corresponding probabilities. 

\begin{figure}
\begin{center}
\newcommand{\sfw}{\figwidth}
\subfigure[]{\includegraphics[width=\sfw]{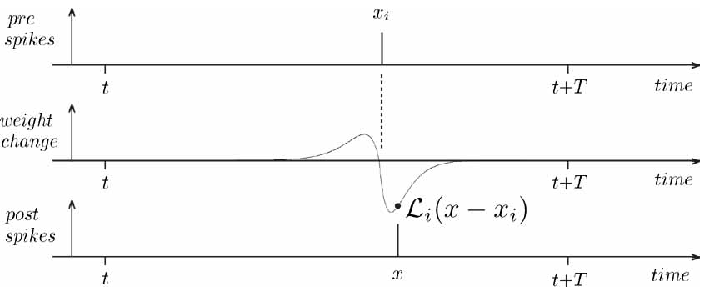}}
\subfigure[]{\includegraphics[width=\sfw]{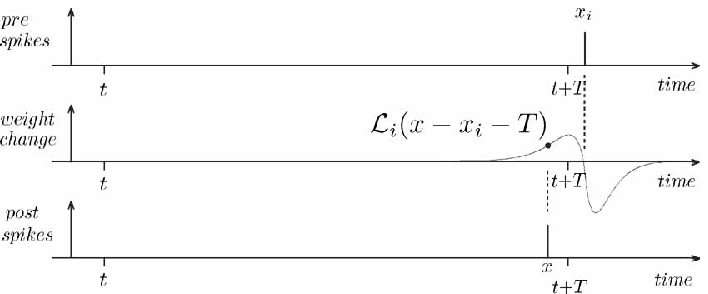}}
\caption{Changes in weight due to pairing of presynaptic and
  postsynaptic spikes. (a) Pairing of a postsynaptic spike at time
  $(x,t)$ and presynaptic spike by cell $i$ at time $(x_i,t)$ causes
  a change $\L_i(x-x_i)$ in weight $w_i$. (b) For $x$ within $\tau_L$ 
of
  a period boundary, we must include pairing with presynaptic spikes in
  the neighboring period. Pairing of a postsynaptic spike at time
  $(x,t)$ and presynaptic spike by cell $i$ at time $(x_i,t+T)$
  cause a change $\L_i(x-x_i-T)$ in weight $w_i$.}
\label{figL}
\end{center}
\end{figure}

The nonassociative change in $w_i(t)$ due to the
single presynaptic spike at $(x_i,t)$ is $\alpha_i$. For the associative
change due to presynaptic and postsynaptic spike pairs, consider the
effect of a single postsynaptic spike at $(x,t)$. The pair consisting of this
spike and the presynaptic spike at $(x_i,t)$ causes a change
$\L_i(x-x_i)$ in $w_i$. To account for pairs which straddle a period boundary, we also 
include
the pairing with presynaptic spikes at $(x_i,t-T)$ and $(x_i,t+T)$,
for a total change of
\begin{equation}
\label{threeL}
\L_i(x-x_i-T) + \L_i(x-x_i) + \L_i(x-x_i+T).
\end{equation}  
For our intended biological application, where $\tau_L \ll T$, at most one
of the above terms is non-negligible, but all must be included to
properly handle cases where $x-x_i$ is within $\tau_L$ of $T$ or $-T$
(Fig. \ref{figL}).
Finally, $\tau_L \ll T$ allows us to approximate
Eq. \eqref{threeL} by
\begin{equation}
\label{singlepost}
\sum_{n=-\infty}^{\infty} \L_i(x-x_i-nT)=\Lo_i(x-x_i),
\end{equation}
where $\Lo_i(s)=\sum_{n=-\infty}^{\infty} \L_i(s-nT)$ is the 
periodization
of $\L_i$ with period $T$.

Quantity \eqref{singlepost} is the change in weight $w_i(t)$ due to a single
postsynaptic spike at $(x,t)$. A postsynaptic spike between $t$ and
$t+T$ occurs with a probability density $\Tinv f(U(x,t))$ in $x$, with the 
probability of zero postsynaptic spikes being $1-\Tintx f(U(x,t))$. 
Hence the change in $w_i$ due to postsynaptic spikes between $t$ and 
$t+T$ is $\Lo_i(x)$ with density $\Tinv f(U(x,t))$ in $x$, and $0$ with 
probability $\quad 1-\Tintx f(U(x,t))$.
The total change in $w_i(t)$ due to both nonassociative and
associative learning is therefore
\begin{equation}
\label{deltawiU}
\tri w_i(t)=\begin{cases} \alpha_i+\Lo_i(x), \text{ density } f(U(x,t)) \\ 
\alpha_i, \text{ probability } 1-\Tintx f(U(x,t)).
\end{cases}    
\end{equation}

\begin{figure}
\begin{center}
\newcommand{\sfw}{\figwidth}
\subfigure[]{\includegraphics[width=\sfw]{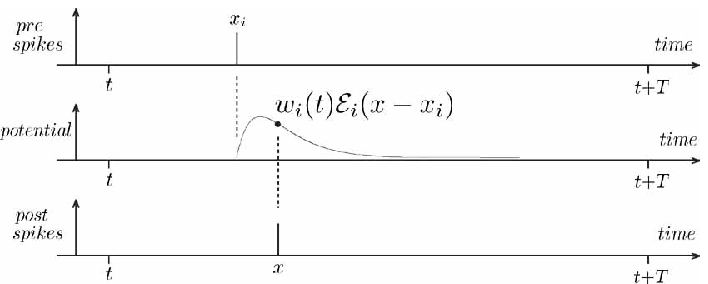}}
\subfigure[]{\includegraphics[width=\sfw]{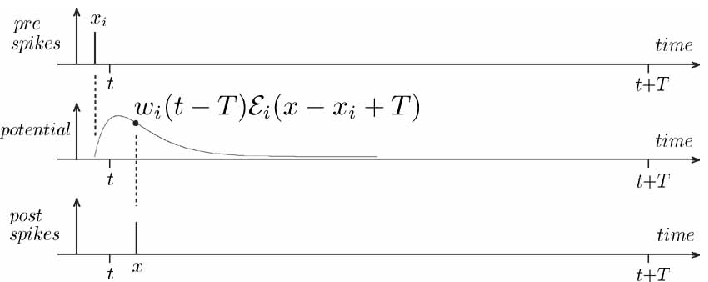}}
\caption{Postsynaptic potential due to presynaptic spikes. (a)
  Potential at time $(x,t)$ due to presynaptic spike by cell $i$ at
  time $(x_i,t)$ is $w_i(t)\E_i(x-x_i)$. (b) For $x$ within $\tau_E$ 
of
  $0$, we must include the potential due to presynaptic spikes in the
  preceding period. The potential at time $(x,t)$ due to the
  presynaptic spike by cell $i$ at time $(x_i,t-T)$ is
  {$w_i(t-T)\E_i(x-x_i+T)$.}}
\label{figE}
\end{center}
\end{figure}

We now compute the non-noisy component of the postsynaptic potential,  $U(x,t)$. The contribution
to $U(x,t)$ from the presynaptic spike by cell $i$ at time $(x_i,t-nT)$
is $w_i(t+nT)\E_i(x-x_i+nT)$. For $\tau_E \ll T$ this quantity is
non-negligible for at most one value of $n$, either the current period ($n=0$) or the previous period ($n=-1$). But to handle edge
effects (Fig. \ref{figE}) we must include both, for a total contribution of
\begin{equation}
\label{twoE}
w_i(t-T)\E_i(x-x_i-T) + w_i(t)\E_i(x-x_i).
\end{equation}
The slow learning rate assumption allows us to approximate quantity \eqref{twoE} by
\begin{equation}
\label{twoEonew}
w_i(t)\bigl[\E_i(x-x_i-T) + \E_i(x-x_i) \bigr].
\end{equation}
Finally, $\tau_E \ll T$ allows us to approximate quantity
\eqref{twoEonew} by
\begin{equation}
\label{singlepre}
w_i(t) \sum_{n=-\infty}^{\infty} \E_i(x-x_i-nT)=w_i(t) \Eo_i(x-x_i),
\end{equation}
where $\Eo_i(s)=\sum_{n=-\infty}^{\infty} \E_i(s-nT)$ is the 
periodization
of $\E_i$ with period $T$.

Quantity \eqref{singlepre} is the contribution to $U(x,t)$ from cell
$i$. The total postsynaptic potential is the summed contribution from
all presynaptic cells, plus the repeated sensory input:
\begin{equation}
\label{Uxt}
 U(x,t) = \phi(x)+\sum_{j=1}^N w_j(t)\Eo_j(x-x_j) 
\end{equation} 
Define $\tif$ by
\begin{eqnarray}
\tif(x,w(t)) & = & f(\phi(x)+\sum_{j=1}^N w_j(t)\Eo_j(x)). 
\end{eqnarray}
Then from Eq. \eqref{deltawiU} and Eq. \eqref{Uxt} we have
\begin{equation}
\label{deltawi}
\tri w_i(t)=\begin{cases} \alpha_i+\Lo_i(x),  \text{ density } 
\Tinv\tif(x,w_1(t),\ldots,w_N(t)) \\ \alpha_i,  \text{ probability } \\
\hfill 1-\Tintx\tif(x,w_1(t),\ldots,w_N(t)).
\end{cases}    
\end{equation}

Eq. \eqref{deltawi} defines the random walk for the weight dynamics.  
It is discrete time (steps occur only at $t=nT$, $n\in \mathbb{Z}$), 
continuous space (steps can take a continuum of values), and 
inhomogenous (step probabilities depend on position). 

The common periodicity of the functions $\Eo_i$, $\Lo_i$
and $\phi$ is an important feature, allowing the systematic use of
Fourier techniques.

\section{One Weight}\label{sec.one}

To illustrate the technique in the simplest possible setting, we first examine the case of a single weight. If there is only one weight, $w_1(t)$, then without loss of 
generality we may take $x_1=0$, by translating $\phi$ if necessary. 
Writing $w(t)$, $\alpha$, $\Lo$ and $\Eo$ for $w_1(t)$, $\alpha_1$, $\Lo_1$ and $\Eo_1$, the random walk Eq. \eqref{deltawi} for the weight dynamics becomes
\begin{equation}
\label{deltawone}
\tri w(t)=\begin{cases} \alpha+\Lo(x), \text{ density } 
\Tinv\tif(x,w(t)) \\ \alpha,  \text{ probability } 1-\Tintx\tif(x,w(t)),
\end{cases}    
\end{equation}
where
\begin{eqnarray}
\tif(x,w(t)) & = & f(\phi(x)+w(t)\Eo(x)). \notag
\end{eqnarray}

From the random walk for the weight dynamics we derive the moments of the equilibrium weight distribution in three steps. First we write the master equation for the time evolution of the probability distribution of the weight, and the corresponding functional equation for the equilibrium (stationary) distribution. Taking the Fourier transform yields a differential equation for the Fourier transform of the equilibrium distribution. Taylor expansion of this 
equation yields recurrence relations for the moments. 

Notice that the set of step values in the walk \eqref{deltawone} is independent of $w$; hence the equilibrium distribution $P(w)$ must satisfy Eq. \eqref{eqbmspecial}. From the step values and step probabilities in Eq. \eqref{deltawone} we have
\begin{eqnarray}
\label{eqbmone}
P(w)=\Bigl[1-\Tintx\tif(x,w-\alpha)\Bigr]P(w-\alpha) \notag \\
+ \Tintx \tif(x,w-(\alpha+\Lo(x)))P(w-(\alpha+\Lo(x))). 
\end{eqnarray}
Taking the Fourier transform $\int dw\, e^{ikw}$ on both sides, 
changing variables and rearranging yields
\begin{eqnarray}
\label{phattif}
\Ph(k)[1-e^{ik\alpha}]=\Tintx [e^{ik(\alpha+\Lo(x)}-e^{ik\alpha}] \notag \\
\times \int dw' \, e^{ikw'} \tif(x,w')P(w').
\end{eqnarray}
A physiologically plausible spike output function $f$ would take the form of a smooth, monotonically increasing sigmoid, but for maximal simplicity we assume $f$ is piecewise 
linear:
\begin{equation}
\label{fpwlin}
f(u)=\begin{cases} 0, & u < -V-\theta \\ 
\TTinv(1+\frac{u-\theta}{V}), & -V-\theta \le u \le V-\theta \\
\Tinv, & u>V-\theta
\end{cases}    
\end{equation}
so that $\tif$ is given by
\begin{equation}
\label{tifone}
\tif(x,w)=\begin{cases} 0, & U(x) < -V-\theta \\ 
\TTinv(1+\frac{U(x)-\theta}{V}), & -V-\theta \le U(x) \le V-\theta \\
\Tinv, & U(x) >V-\theta
\end{cases}    
\end{equation}
with $U(x)=\phi(x)-\theta +w\,\Eo(x)$. 

We further assume that the equilibrium weight distribution $P(w)$ is zero 
or negligible for $w$ such that $U(x)< -V-\theta$ or $U(x)> 
V-\theta$. This is a \emph{confinement condition} on the equilibrium 
postsynaptic potential $U(x)$, and will be justified later. Note that the confinement condition helps justify the piecewise linear assumption on $f$, since the more ``confined'' the postsynaptic potential $U(x)$, the better our piecewise linear $f$ approximates a smooth sigmoid in the region where $U(x)$ is concentrated. If the 
confinement condition holds, then in Eq. \eqref{phattif} we may 
replace $\tif(x,w')$ under the integral by the following linear function of $w$: 
\begin{equation}
\label{ }
\TTinv (1+\frac{\phi(x)-\theta +w\,\Eo(x)}{V}).  \notag
\end{equation}
Using $\int dw\, e^{ikw} w P(w)=\Ph'(k)$, we then obtain 
\begin{eqnarray}
\label{odeph}
\Ph(k)\Bigl[1-e^{ik\alpha}-\Tintx \half 
(1+\frac{\phi(x)-\theta}{V})\eta(x)\Bigr] \notag \\
= \iinv\Ph'(k) \Tintx \half \frac{\Eo(x)}{V} 
\eta(x),
\end{eqnarray}
where $\eta(x)=e^{ik(\alpha+\Lo(x)}-e^{ik\alpha}$.
By Eq. \eqref{derivsofchar}, the moments of $P(w)$ are (up to powers of $i$) just the 
derivatives of $\Ph(k)$ at $k=0$; since those derivatives are 
implicitly constrained by Eq. \eqref{odeph}, the moments of $P(w)$ 
are constrained by Eq. \eqref{odeph}. Specifically, the Taylor 
expansion of Eq. \eqref{odeph} around $k=0$ will yield a hierarchy of 
recurrence relations for the derivatives of $\Ph(k)$, and hence for the moments of $P(w)$. The Taylor expansions of the exponentials are 
\begin{eqnarray*}
\label{ }
e^{ik\alpha} & = & \suminfn \infact \alpha^n k^n,  \\
e^{ik(\alpha+\Lo(x))} & = & \suminfn \infact (\alpha+\Lo(x))^n k^n.
\end{eqnarray*}

For the expansion of the characterisitic function $\Ph(k)$ we expand 
the exponential in the definition of $\Ph(k)$ and invert the order of 
summation and integration:
\begin{eqnarray*}
\Ph(k) & = & \int dw\, e^{ikw} P(w) \notag \\
 & = & \suminfm \imfact k^m \int dw\, w^m P(w) \notag \\
& = & \suminfm \imfact \m{w^m} k^m.
\end{eqnarray*}

From this it follows that 
\begin{eqnarray*}
\iinv \Ph'(k) & = & \iinv  \suminfm \imfact \m{w^m} k^{m-1} m 
\notag \\
& = & \suminfm \imfact \m{w^{m+1}} k^m.
\end{eqnarray*}

By substituting these expansions into Eq. \eqref{odeph} and equating 
coefficients of $k^\mu$ on both sides, we obtain the following 
relations:
\begin{eqnarray}
\label{recurrone}
\sum_{m=0}^\mu \binom{\mu}{m}\bigl[\gamp_{\mu-m} \m{w^m} &-&   
\game_{\mu-m} \m{w^{m+1}} \bigr] =0 , \\
  \mu &=& 0,1,2,\ldots \notag 
\end{eqnarray}
where for brevity we have defined 
\begin{eqnarray*}
\gamp_n & = & \delta_{n,0} -\alpha^n \\
&& -\Tintx \half 
(1+\frac{\phi(x)-\theta}{V})((\alpha+\Lo(x))^n-\alpha^n), \notag \\
\game_n & = & \Tintx \half \frac{\Eo(x)}{V} 
(\alpha+\Lo(x))^n-\alpha^n) .
\end{eqnarray*}


The relations \eqref{recurrone} are lower triangular\footnote{One could also derive moment equations via the more direct route of Taylor expanding, in $w$, the equilibrium condition \eqref{eqbmone} for $P(w)$; but the resulting moment equations are not triangular. In fact they are fully coupled (each equation involving all moments, in general) and hence not readily solvable.}, and hence are easily rearranged to yield 
explicit recurrence relations for the moments in terms of moments of 
lower degree only:
\begin{eqnarray}
\label{recurronewmu}
\m{w^\mu} = -\frac{\gamp_\mu}{\mu\game_1} &-& \frac{1}{\mu\game_1} 
\sum_{m=1}^{\mu-1}  \m{w^m}  \, \psi_{\mu,m}, \\
  \mu &=& 1,2,\ldots  \notag 
\end{eqnarray}
where 
\begin{equation}
\label{ }
\psi_{\mu,m} = \binom{\mu}{m}\gamp_{\mu-m} - \binom{\mu}{m-1} \game_{\mu-m+1}. \notag
\end{equation}

We may now compute the central moments 
$M_k=\m{(w-\m{w})^k}$, by expressing 
$\m{w^n}$ in terms of the $\{M_k\}$:
\begin{eqnarray}
\label{Mk}
\m{w^n} &=& \m{(w-\m{w}+\m{w})^n} \notag \\
&=& \sum_{k=0}^n \binom{n}{k} M_k \m{w}^{n-k}.
\end{eqnarray}
Substituting into Eq. \eqref{recurronewmu} and rearranging yields
\begin{eqnarray}
\label{recurroneMmu}
 M_\mu & = & -(\gamfrac)^\mu -\frac{\gamp_\mu}{\mu \game_1} - 
\frac{1}{\mu \game_1} \sum_{m=1}^{\mu-1} (\gamfrac)^m  \, \psi_{\mu,m}     \notag      \\
& +  &  \sum_{k=2}^{\mu-1} M_k \Bigl\{-\binom{\mu}{k} 
(\gamfrac)^{\mu-k} \notag \\
&-& \frac{1}{\mu 
\game_1}\sum_{m=k}^{\mu-1}\binom{m}{k}(\gamfrac)^{m-k} \, \psi_{\mu,m} \Bigr\}.
\end{eqnarray}

For $\mu=2,3,4$ we obtain
\begin{eqnarray}
\label{firstfewM}
M_2 &=& -\frac{1}{2}\frac{\gamp_2}{\game_1} 
+ \frac{1}{2}\frac{\gamp_1\game_2}{(\game_1)^2},  \notag \\
M_3 &=& -\frac{1}{3}\frac{\gamp_3}{\game_1} 
+ \frac{1}{3}\frac{\gamp_1\game_3}{(\game_1)^2} 
+ \frac{1}{2}\frac{\gamp_2\game_2}{(\game_1)^2} 
- \frac{1}{2}\frac{\gamp_1(\game_2)^2}{(\game_1)^3},  \notag \\
M_4 &=& \frac{3}{4}\frac{(\gamp_2)^2}{(\game_1)^2}
-\frac{3}{2}\frac{\gamp_1\gamp_2\game_2}{(\game_1)^3}
+\frac{3}{4}\frac{(\gamp_1)^2(\game_1)^2}{(\game_1)^4} - \frac{1}{4}\frac{\gamp_4}{\game_1} \notag \\
&& + \frac{1}{4}\frac{\gamp_1\game_4}{(\game_1)^2} 
 + \frac{1}{2}\frac{\gamp_2\game_3}{(\game_1)^2}
+ \frac{1}{2}\frac{\gamp_3\game_2}{(\game_1)^2} \notag \\
&& - \frac{\gamp_1\game_2\game_3}{(\game_1)^3}
- \frac{3}{4}\frac{\gamp_2(\game_2)^2}{(\game_1)^3}
+ \frac{3}{4}\frac{\gamp_1(\game_2)^3}{(\game_1)^4}.
\end{eqnarray}

We can see from $M_3$ alone that in general the equilibrium weight 
distribution is not Gaussian. For generic PSP $\E$ and learning rule 
$\L$ there are no polynomial relations amongst the coefficients 
$\game_n$ and $\gamp_n$, hence $M_3$ is generically nonzero. 


To determine the dependence of the moments on step size, we multiply both $\alpha$ and $\L$, and hence the steps of the random walk, by a scalar  
$\lambda$. The 
coefficients $\game_n$ and $\gamp_n$ are then both $O(\lambda^n)$, and 
substitution into Eq. \eqref{firstfewM} yields
\begin{eqnarray*}
M_2 & = & O(\lambda), \\
M_3 & = & O(\lambda^2), \\
M_4 & = & 3M_2^2 + O(\lambda^3).
\end{eqnarray*}

Hence as $\lambda \to 0$ the skew and kurtosis approach Gaussian 
values:

\begin{eqnarray*}
\text{skew} & = & \frac{M_3}{M_2^{3/2}}  =  O(\lambda^{\frac{1}{2}})  
\to  0,\\
\text{kurtosis} &  = & \frac{M_4}{M_2^2}  =  3 + O(\lambda)  \to  3.
\end{eqnarray*}

\section{Multiple Weights}\label{sec.mult}

We now apply the technique to the case of multiple weights $w_i$, 
$i=1,2,\ldots,N$. The algebra is more complicated, but the structure 
of the derivation is identical to the single weight case. For 
notational compactness we introduce vector notation:
\begin{eqnarray*}
w(t) & = & \begin{pmatrix} w_1(t) \\ \vdots \\ w_N(t) 
\end{pmatrix} , \quad \alphab  = \begin{pmatrix} \alpha_1 \\ \vdots 
\\ \alpha_N \end{pmatrix}, \\
\Eo(x) & = & \begin{pmatrix} \Eo_1(x-x_1) \\ \vdots \\ \Eo_N(x-x_N) 
\end{pmatrix} , \quad \Lob(x)  =  \begin{pmatrix} \Lo_1(x-x_1) \\ 
\vdots \\ \Lo_N(x-x_N) \end{pmatrix}.
\end{eqnarray*}
The random walk for the weight vector $w(t)$ takes place in 
$\mathbb{R^N}$, with the walk for each component $w_i(t)$ given by 
Eq. \eqref{deltawone}. In vector notation the walk for $w(t)$ is then
\begin{equation}
\label{deltawvec}
\tri w(t)=\begin{cases} \alpha+\Lo(x),  \text{ density }
\Tinv\tif(x,w(t)) \\ \alpha,  \text{ probability }  1-\Tintx\tif(x,w(t)),
\end{cases}    
\end{equation}
where  
\begin{eqnarray*}
\tif(x,w(t)) & = & f(\phi(x)+w(t)\cdot\Eo(x)) .
\end{eqnarray*}
and $\cdot$ indicates the vector dot product.
Again, the step sizes are independent of position, so the equilibrium condition Eq. \eqref{eqbmspecial} applies. We have
\begin{eqnarray}
\label{eqbmvec}
P(w)=\Bigl[1-\Tintx\tif(x,w-\alpha)\Bigr]P(w-\alpha) \notag \\
+ \Tintx \tif(x,w-(\alpha+\Lo(x)))P(w-(\alpha+\Lo(x))).
\end{eqnarray}
As before, we take the (now $n$-dimensional) Fourier transform on both sides. Applying $\int dw\, 
e^{ik\cdot w}$, changing variables and rearranging 
yields
\begin{equation}
\label{phattif}
\Ph(k)[1-e^{ik\cdot\alpha}]=\Tintx 
\eta(x) \int dw' \, e^{ik\cdot 
w'} \tif(x,w')P(w'),
\end{equation}
where
\begin{equation}
\eta(x)=e^{ik\cdot(\alpha+\Lo(x))}-e^{ik\cdot\alpha}.
\end{equation}
We now assume the postsynaptic gain function $f$ is piecewise linear 
and given by Eq. \eqref{fpwlin}, hence $\tif$ is given by Eq. 
\eqref{tifone}, with $U(x)=\phi(x)-\theta +w\cdot\Eo(x)$. And as 
before, we assume $P(w)$ is negligible for $w$ such that $U(x)< 
-V-\theta$ or $U(x)> V-\theta$, a confinement condition on $P(w)$, 
which will be justified later. Then we may replace $\tif(x,w')$ under 
the integral by the linear function of $w$ 
\begin{equation}
\label{ }
\TTinv (1+\frac{\phi(x)-\theta +w\cdot\Eo(x)}{V}).  \notag
\end{equation}
Using $\int dw\, e^{ik\cdot w} w_j P(w)=\frac{1}{i} 
\del{\Ph(k)}{k_j}$, we obtain the following first-order PDE for $\Ph(k)$: 
\begin{eqnarray}
\label{pdephvec}
\Ph(k)\Bigl[1-e^{ik\cdot\alpha}-\Tintx \half 
(1+\frac{\phi(x)-\theta}{V})\eta(x)\Bigr] 
\notag \\
=  \sum_{j=1}^{N} \frac{1}{i} \del{\Ph(k)}{k_j} \Tintx \half 
\frac{\Eo_j(x-x_j)}{V} \eta(x).
\end{eqnarray}
Taylor expansion of both sides of this equation around $k=0$ 
will yield recurrence relations for the 
moments of $w$. The Taylor expansion of a function $g$ on 
$\mathbb{R^N}$ is given by 
\begin{eqnarray}
\label{taylorRN}
g(k_1,\ldots,k_N)&=&  \suminfn\frac{1}{n!} \sum_{\substack{0\le s_l 
\le n\\ \sum s_l=n}}  \binom{n}{s_1\, \cdots \, s_N} \notag \\
& \times & \frac{\partial^n 
g(z_1,\ldots,z_N)}{\partial^{s_1}_{z_1}\cdots 
\partial^{s_N}_{z_N}} \Bigr|_{z=0} \prod_l k_l^{s_l}.
\end{eqnarray}
The expansions of the complex exponentials in Eq. \eqref{pdephvec} 
are thus
\begin{eqnarray}
\label{exponexpvec}
e^{ik\cdot\alpha} & = & \suminfn \sum_s \infact \binom{n}{s} \prod_l 
\alpha_l^{s_l} \prod_l k_l^{s_l},  \notag \\
e^{ik\cdot(\alpha+\Lo(x))} & = & \suminfn \sum_s \infact \binom{n}{s} \notag \\
&& \times \prod_l (\alpha+\Lo(x))_l^{s_l} \prod_l k_l^{s_l},
\end{eqnarray}
where in the sums on the right, $s=(s_{1} s_{2} \ldots s_{N})^{T}$ with each $s_{i}$ a nonnegative integer and $\sum_{i=1}^{N}s_{i}=n$. For brevity we write $\binom{n}{s}$ for the multinomial coefficient in Eq. \eqref{taylorRN}.

As before, for the expansion of the characterisitic function $\Ph(k)$ 
we expand the exponential in the definition of $\Ph(k)$ and invert 
the order of summation and integration:
\begin{eqnarray}
\label{phexpvec}
\Ph(k) & = & \int dw\, e^{ik\cdot w} P(w) \notag \\
 & = & \suminfm \sum_r \imfact \binom{m}{r} \bigl[\int dw\, P(w) 
\prod_l w_l^{r_l} \bigr] \prod_l k_l^{r_l}  \notag \\
 & = & \suminfm \sum_r \imfact \binom{m}{r} \m{w_1^{r_1} w_2^{r_2} 
\ldots w_N^{r_N}}  \prod_l k_l^{r_l}, 
\end{eqnarray}
where $r=(r_{1} r_{2} \ldots r_{N})^{T}$ with each $r_{i}$ a nonnegative integer and $\sum_{i=1}^{N}r_{i}=m$. From this expansion of $\Ph(k)$ it follows that 
\begin{equation} 
\label{dphexpvecalmost}
\frac{1}{i} \del{\Ph(k)}{k_j}  =  \suminfm \sum_r  \imfact \binom{m}{r} \m{w_1^{r_1} \cdots w_N^{r_N}}  r_j \prod_l k_l^{r_l-\delta_{lj}} .
\end{equation}
Using the combinatorial identity
\begin{equation}
\label{ }
\frac{i^{m-1}}{m!} \binom{m}{r} r_j = \frac{i^{m-1}}{(m-1)!} 
\binom{m-1}{r_1 \cdots r_j-1 \cdots r_N}, \notag
\end{equation}
we may reindex Eq. \eqref{dphexpvecalmost} to yield
\begin{eqnarray} 
\label{dphexpvec}
\frac{1}{i} \del{\Ph(k)}{k_j}  &=&  \suminfm \sum_r  \imfact 
\binom{m}{r} \notag \\
&& \times \m{w_1^{r_1} \cdots w_j^{r_j+1}  \cdots w_N^{r_N}}  
\prod_l k_l^{r_l} .
\end{eqnarray}
When the expansions Eqs. \eqref{exponexpvec}, \eqref{phexpvec}, and 
\eqref{dphexpvec} are substituted into Eq. \eqref{pdephvec}, equating 
the coefficients of $\prod_l k_l^{q_l}$ on both sides yields  
\begin{eqnarray}
\label{recurrmuvecalmost}
 \frac{1}{\mu !} \binom{\mu}{q} && \m{w_1^{q_1} w_2^{q_2} \cdots 
w_N^{q_N}}  \notag \\
 && =  \sum_{r+s=q} \frac{1}{n!m!} \binom{m}{r} \binom{n}{s} 
\Bigl[  \gamp_s \m{w_1^{r_1} w_2^{r_2} \cdots w_N^{r_N}}  \notag \\
&& + \sum_{j=1}^N  \gamej_s \m{w_1^{r_1} \cdots w_j^{r_j+1}  \cdots 
w_N^{r_N}}  \Bigr], 
\end{eqnarray}
where 
\begin{eqnarray*}
\gamp_s  &=& \Tintx \half (1+\frac{\phi(x)-\theta}{V})(\prod_l(\alpha+\Lo(x))_l^{s_l}-\prod_l  \alpha_l^{s_l})  \notag \\
&+&  \prod_l \alpha_l^{s_l}, \notag \\
\gamej_s &=&  \Tintx  \half \frac{\Eo_j(x-x_j)}{V} (\prod_l(\alpha+\Lo(x))_l^{s_l}-\prod_l \alpha_l^{s_l}),
\end{eqnarray*}
and $q=(q_{1} q_{2} \ldots q_{N})^{T}$, each $q_{i}$ a nonnegative integer, with $\sum_{i=1}^{N}q_{i}=\mu$. A slight simplification follows from $\gamp_{0}=1$ and $\game_{0}=0$: the quantity on the left side of Eq. \eqref{recurrmuvecalmost} is 
cancelled by the term on the right side with $s=0$ and $r=q$. The 
resulting recurrence relations are
\begin{eqnarray}
\label{recurrmuvec}
0 &=& \sum_{\substack{r+s=q \\ m<\mu}} \frac{1}{n!m!} \binom{m}{r} 
\binom{n}{s} \Bigl[  \gamp_s \m{w_1^{r_1} w_2^{r_2} \cdots 
w_N^{r_N}}  \notag \\
&& +  \sum_{j=1}^N  \gamej \m{w_1^{r_1} \cdots 
w_j^{r_j+1}  \cdots w_N^{r_N}}  \Bigr],  \\
0 & \le & q_{i} \le  \mu, \quad \sum_{i=1}^{N}q_{i}=\mu, \quad \mu  = 1,2,\ldots   \notag
\end{eqnarray}
For each choice of $q$ we obtain a single linear equation involving moments of total order at most $\mu=\sum_{i}q_{i}$. Regarding the moments of total order $\mu$ as unknowns, to be solved for in terms of moments of total order less than $\mu$, we have a linear system with the same number of equations as unknowns. The coefficient matrix of this system involves the quantities $\gamp_{\cdot}$ and $\game_{\cdot}$. For generic $\E$ and $\L$ there are no polynomial relations amongst these quantities; hence the determinant of the coefficient matrix is generically nonzero, and the system can be inverted to give the moments of total order $\mu$ in terms of $\gamp_{\cdot}$, $\game_{\cdot}$, and the moments of total order less than $\mu$. The complete moment hierarchy can thus be obtained: first moments of total order $1$, then moments of total order $2$, and so on.

\subsection{Equilibrium Mean}

For $\mu=1$ we must have $q_{j}=\delta_{ij}$ for some $i$. Since in Eq. \eqref{recurrmuvec} only terms with $m<\mu$ appear, and $m=\sum_{j}r_{j}$, the only possibility for $r$ is $r=0$, and then $s_{j}=q_{j}=\delta_{ij}$. The recurrence relation Eq. \eqref{recurrmuvec} then becomes
\begin{eqnarray}
\label{meanwveccomp}
0 &=&  \alpha_i + \Tintx \half (1+\frac{\phi(x)-\theta}{V}) \Lo_i(x-x_i) \notag \\ 
&+& \sum_{j=1}^N \m{w_j} \Tintx \frac{1}{2V} \Eo_j(x-x_j)\Lo_i(x-x_i).
\end{eqnarray}
Allowing $i$ to vary over all possible values $1,2,\ldots,N$, we have $N$ linear equations in the $N$ unknowns $w_{i}$, which can be written in vector form as
\begin{equation} 
\label{Cwd}
C\m{w} = d,
\end{equation}
with the matrix $C$ and vector $d$ given by
\begin{eqnarray}
C_{ij} & = & - \frac{1}{2V} \Tintx \Eo_j(x-x_j) \Lo_i(x-x_i), \label{C}\\
d_i & = &  \alpha_i  \Tintx \half (1+\frac{\phi(x)-\theta}{V}) 
\Lo_i(x-x_i).  \label{d} 
\end{eqnarray}
The overall minus sign in the definition of $C$ is for later convenience. For generic $\E$ and $\L$ the matrix $C$ is invertible, and we have $\m{w}=C^{-1}d$. The physical meaning of this relation can be illuminated by rewriting Eq. \eqref{meanwveccomp} as follows:
\begin{eqnarray*}
0 &=& \alpha_i \notag \\
&+& \Tintx (1+\frac{\phi(x)-\theta+\sum_{j=1}^N 
\m{w_j} \Eo_j(x-x_j)}{V}) \Lo_i(x-x_i)  \\
 & = & \alpha_i + \Tintx \m{f}(x) \Lo_i(x-x_i),
\end{eqnarray*}
where $\m{f}(x)$ we define to be the value of $f(x)$ when $w=\m{w}$. Now add and subtract $\alpha_{i}\Tintx \m{f}(x)$ to obtain
\begin{eqnarray}
\label{meanfmeanw}
0 &=& \bigl[1-\Tintx  \m{f}(x) \bigr] \alpha_i \notag \\
&+& \Tintx 
\m{f}(x) (\alpha_i +\Lo_i(x-x_i)) \notag \\
& = & \m{\tri w_i}. 
\end{eqnarray}
We find that the equilibrium mean weight vector $\m{w}$ is that for which the mean weight change is zero for all weights. 
This condition is obvious on independent grounds, and could have been used to calculate $\m{w}$ directly, without recourse to the moment hierarchy relations. But for moments of total order $2$ or higher, transparent conditions such as this are not available; in that case we have no choice but to solve Eq. \eqref{recurrmuvec}.
 
Given the equilibrium mean weights $\m{w}$, we can calculate the equilibrium mean postsynaptic potential $\m{U}(x)$ via
\begin{eqnarray*}
\label{ }
\m{U}(x) & = & \phi(x) + \Eo(x) \cdot \m{w}  \notag \\
 & = & \phi(x) + \Eo(x) \cdot C^{-1}d,
\end{eqnarray*}
provided $C$ is invertible.

\subsection{Equilibrium Variance}

We now take $\mu=2$ and $q_{k}=\delta_{ik}+\delta_{jk}$ in Eq. \eqref{recurrmuvec}. After some simplification, using $C$, $d$, $\m{f}$ and $\m{w}$ from above, we obtain
\begin{eqnarray*}
\label{ }
&0& = - \sum_{k=1}^N C_{jk} \m{w_k w_i}  -  \sum_{k=1}^N C_{ik} 
\m{w_k w_j} \notag \\
& - & \m{w_i} d_j - \m{w_j} d_i  + \Tintx \m{f}(x) \\
& \times & \bigl[ (\alpha_i +\Lo_i(x-x_i))(\alpha_j +\Lo_j(x-x_j)) -\alpha_i 
\alpha_j \bigr]. \notag
\end{eqnarray*}
This can be rearranged to give 
\begin{eqnarray*}
&&\sum_{k=1}^N C_{jk} \m{w_k w_i} + \sum_{k=1}^N C_{ik} \m{w_k 
w_j} \notag \\
& - & \m{w_i} \sum_{k=1}^N C_{jk} \m{w_k} - \m{w_j} 
\sum_{k=1}^N C_{ik} \m{w_k} \notag \\
& = & \Tintx \m{f}(x) \bigl[ 
(\alpha_i +\Lo_i(x-x_i))(\alpha_j +\Lo_j(x-x_j)) \bigr]  \\
&& + \Bigl[ 
1  - \Tintx  \m{f}(x) \Bigr] \alpha_i \alpha_j \notag \\
 & = & - \m{\tri w_i \tri w_j}.
\end{eqnarray*}
In vector form this becomes
\begin{eqnarray}
\label{lyapvarwalmost}
C(\m{w w^{T}}-\m{w}\m{w}^T) &+& (\m{w^2}-\m{w}\m{w}^T)C^T \notag \\
&& = \m{\tri w \tri w^{T}}.
\end{eqnarray}
The covariance of a vector random variable $v$ is $\var v = \m{v v^{T}}-\m{v}\m{v}^T$. Equation \eqref{lyapvarwalmost} then takes the compact form 
\begin{equation}
\label{lyapvarw}
C (\var w) + (\var w) C^T =  \var \tri w,
\end{equation}
where we have used the equilibrium mean condition $\m{\tri w}=0$ on the right side. Equation \eqref{lyapvarw} is a Lyapunov equation \cite{bhatiarosenthal} for $\var w$, giving the equilibrium weight covariance in terms of $C$ (which depends on $\E$ and $\L$) and $\var \tri w$ (which depends on $\m{f}$, $\alpha$, and $\L$). Both $C$ and $\var \tri w$ can be calculated from the parameters of the system, and then the equilibrium covariance $\var w$, if it exists, must satisfy Eq. \eqref{lyapvarw}. 

A theorem of Ostrowski and Schneider \cite{ostrowskischneider,bhatiarosenthal} gives conditions for existence and uniqueness of solutions to Lyapunov equations. If $S$ is symmetric positive definite and $A$ and $-A$ have no common eigenvalues, then the Lyapunov equation $AH + HA^T = S$ has a unique solution $H$. Furthermore, $H$ is symmetric, and has the same \emph{inertia} (number of eigenvalues with positive, zero, or negative real part) as $A$. 

Since $\var \tri w$ is necessarily symmetric positive definite, the theorem says that a symmetric solution $\var w$ to Eq. \eqref{lyapvarw} exists uniquely provided $C$ and $-C$ have no common eigenvalues, and $\var w$ is positive definite if and only if all eigenvalues of $C$ have positive real part. 

The condition that $C$ and $-C$ have no common eigenvalues is true for generic $C$ and hence for generic $\E$ and $\L$. The condition that $\var w$ be positive definite is needed in order to interpret $\var w$ as the covariance matrix of a probability distribution; we say $\var w$ is \emph{physical} if it is positive definite. Denoting by $\lambda_n^C$ the $n^\text{th}$ eigenvalue of $C$, we then have the following physicality condition:
\begin{eqnarray}
\label{physicality}
\var w \text{ physical} \iff Re\, \lambda_n^C > 0 \tforall n .
\end{eqnarray}
A theorem of Heinz \cite{heinz, bhatiarosenthal} says that if all eigenvalues of $A$ have positive real part and all eigenvalues of $B$ have negative real part, then the (unique) solution $X$ to the equation $AX-XB=Y$ is given by 
\begin{equation}
\label{heinzx}
X= \int_0^\infty ds \, e^{-sA} Y e^{sB},
\end{equation} 
where the matrix exponentials are defined via Taylor expansions. The assumptions on the eigenvalues of $A$ and $B$ ensure that the integral in Eq. \eqref{heinzx} converges, and one can show by direct substitution that the resulting $X$ satisfies $AX-XB=Y$. If the physicality condition \eqref{physicality} holds, then $C$ and $-C^{T}$ satisfy the conditions for $A$ and $B$ respectively, and we obtain
\begin{equation}
\label{varwintegral}
\var w = \int_0^\infty ds \, e^{-sC} (\var \tri w) e^{-sC^T}. 
\end{equation}
This gives the equilibrium covariance matrix explicitly in terms of system parameters.   

Since the postsynaptic potential $U(x)$ is a deterministic function of the synaptic weight vector $w$, the weight covariance $\var w$ determines the covariance of the postsynaptic potential. From $U(x)=\phi(x)+\Eo(x)w$, we have \begin{equation}
\label{varUxy}
\var(U(x),U(y)) = \Eo(x)^T \, \var w \,\, \Eo(y)
\end{equation}
for any pair of times $x,y$ in the interval $[0,T]$. Of particular interest is the diagonal variance of $U(x)$:
\begin{equation}
\label{varUxx}
\var(U(x),U(x))  = \Eo(x)^T \, \var w \,\, \Eo(x)
\end{equation}
Our derivation of the equilibrium moment hierarchy equations relied on the equilibrium distribution of $U(x)$ being negligible on the ``tails'' of the postsynaptic spike probability function $f$.  We will show in the next section, for the case of homogeneous parameters, that the confinement condition on $U(x)$ can be always be satisfied by adjusting the rates of associative and non-associative learning.  

Note that for a spatially extended psp $\E$, Eq. \eqref{varUxx} implies that the diagonal variance of $U(x)$ depends on the full matrix $\var w$; in other words, it depends not only on the diagonal variances of the synaptic weights $w$, but also on the off-diagonal correlations between different synaptic weights.

\section{Multiple Weights, Homogeneous Parameters}\label{sec.hom}
 
 
For maximal generality in the foregoing analysis, we have allowed the postsynaptic potential functions and spike-timing dependent learning rules to be different for different presynaptic neurons, and have allowed the presynaptic spike times to be arbitrary. Further analytical progress can be made in the case where the system parameters are homogeneous, i.e. the postsynaptic potential functions and spike-timing dependent learning rules are the same for all presynaptic neurons, and the presynaptic spike times are regularly spaced. 

For such parameters it will turn out that the matrix $C$, the coefficient matrix in the Lyapunov equation \eqref{lyapvarw} for $\var w$, has a special form: it is \emph{circulant} \cite{davis}. The matrix $\var \tri w$ on the right side of the Lyaponov equation for $\var w$ is not circulant in general; but it is circulant if the postsynaptic spike probability density $\m{f}(x)$ is independent of $x$. Now it was shown in \cite{williams03} that in the case of homogeneous parameters, if the spacing $\delta$ between presynaptic spike times  is sufficiently small and provided certain other constraints hold, the (mean) equilibrium weight vector has the property that the mean total postsynaptic potential $\m{U}(x)$ is approximately constant\footnote{The present model differs from the model in \cite{williams03} in having a postsynaptic spike probability density instead of a mean postsynaptic spike rate, but the argument is unaffected.}. In that case the mean equilibrium postsynaptic spike density $\m{f}(x)$ is also approximately constant, and the matrix $\var \tri w$ is approximately a circulant matrix $D$. The Lyapunov equation for $\var w$ is then approximately
\begin{equation}
\label{lyapCD}
C \, (\var w) + (\var w) \, C^{T} = D,
\end{equation} 
with solution given by 
\begin{equation}
\label{heinzsolnCD}
\var w = \int_{0}^{\infty} ds \, e^{-sC}D e^{-sC^{T}}. 
\end{equation}
The eigenvalues and eigenvectors of circulant matrices are easily calculated; furthermore, all circulant matrices can be \emph{simulataneously} diagonalized. Simultaneous diagonalization of $C$, $C^{T}$, and $D$ in Eq. \eqref{heinzsolnCD} will yield an explicit solution for $\var w$ in terms of the eigenvectors and eigenvalues of $C$ and $D$, which will themselves be written as explicit functions of the system parameters. 
 
Let $\E(s)$, $\L(s)$, and $\alpha$ denote the common postsynaptic potential function, associative learning rule, and nonassociative learning rule respectively. Let the spike time for presynaptic cell $i$ be $x_{i}=(i-1)\delta$, $i=1,2,\ldots,N$, $\delta=T/N$. We then have
\begin{equation}
\label{Chom}
C_{ij} =  -\frac{1}{2V} \Tintx \Lo(x-x_i) \Eo(x-x_j), 
\end{equation}
and for $\m{f}(x)$ approximately the constant $\m{f}$ we have $\var \tri w \simeq D$, where
\begin{eqnarray*}
\label{}
D_{ij} & = & \alpha^{2} (1-\m{f}) \notag \\
& + & \m{f} \Tintx
(\alpha +\Lo(x-x_i))(\alpha +\Lo(x-x_j)).
\end{eqnarray*}
By periodicity of $\Lo$, this can be simplified to 
\begin{eqnarray}
\label{Dhom}
D_{ij} & = &  \alpha^2 + 2\m{f}\alpha \beta \notag \\
& + & \m{f} \Tintx \Lo(x-x_i) \Lo(x-x_j) ,
\end{eqnarray}
where $\beta = \Tintx \Lo(x)$. 

A matrix $A$ is circulant \cite{davis} if each row of $A$ equals the row above it shifted one entry to the right (and wrapped around at the edges); in other words
\begin{equation}
\label{ }
A_{(i+1) \text{\,mod\,}N,(j+1) \text{\,mod\,}N} = A_{ij}  \tforall i,j.  \notag
\end{equation}
We now show that both $C$ and $D$ are circulant. First, let
$g(x)$ and $h(x)$ be any periodic functions of $x$ with period $T$, and let the $\{x_{i}\}$ be regularly spaced on $[0,T]$ as defined above. Let $A$ be the matrix defined by
\begin{equation}
\label{Aij}
A_{ij} = \int_{0}^{T}dx\, f(x-x_{i}) g(x-x_{j}). 
\end{equation}
Taking $(i,j)$ to $((i+1)\text{\,mod\,}N,(j+1)\text{\,mod\,}N)$ in Eq. \eqref{Aij} shifts the argument of both functions by $-\delta$, and by periodicity this does not change the value of the integral. Hence any matrix of the form \eqref{Aij} is circulant. 

The constant matrices (all of whose entries are the same) are also circulant; and circulant matrices are closed under addition, scalar multiplication, and transposition. Hence by Eq. \eqref{Chom} and \eqref{Dhom}, $C$ and $D$ are both circulant, and so is $C^{T}$.

It is easily shown \cite{davis} that the vectors $u^{(n)}$, $n=1,2,\ldots,N$ with components
\begin{equation}
\label{ungeneral}
u^{(n)}_{l} = e^{2\pi i (l-1)n/N }, \quad k=1,2,\ldots, N
\end{equation}
are a complete set of eigenvectors for any circulant matrix $A$, with corresponding eigenvalue $\lambda_{n}$ given by
\begin{equation}
\label{lambdangeneral}
\lambda_{n} = \sum_{l=1}^{N}A_{jl} e^{2\pi i (j-l) n/N } .
\end{equation}
The expression on the right in Eq. \eqref{lambdangeneral} is independent of $j$ because $A_{jl}$ and the complex exponential both depend only on $(j-l)$ mod $N$. It is easily checked from Eq. \eqref{lambdangeneral} that adding a constant matrix (all entries the same) to a nonzero circulant matrix has no effect on its eigenvalues.

Let $R$ be the unitary matrix whose $n^{th}$ column is the vector $u^{(n)}$, and let $\Lambda$ be the diagonal matrix with entries $\lambda_{n}$. Then  
\begin{equation} 
\label{ }
A  = R \Lambda R^{*}    \notag
\end{equation}
where $R^{*}$ is the complex conjugate transpose of $R$. 


In the present context it will be convenient to define wavenumbers $k_{n}$ so that the argument of the complex exponential in Eq. \eqref{ungeneral} is $i k_{n}x_{l}$; this we can arrange by taking $k_{n}=2 \pi n/T$, $n=1,2,\ldots,N$. From Eq. \eqref{lambdangeneral}, the eigenvalues of $C$ and $D$ are then
\begin{eqnarray*}
\lambda^{C}_{n} & = & - \frac{1}{2V} \sum_{l=1}^{N} e^{ik_{n}(x_{j}-x_{l})} \Tintx \Lo(x-x_j) \Eo(x-x_l), \\
\lambda^{D}_{n} & = & \m{f} \sum_{l=1}^{N} e^{ik_{n}(x_{j}-x_{l})} \Tintx \Lo(x-x_j) \Lo(x-x_l). 
\end{eqnarray*}
By periodicity of $\Eo$ and $\Lo$ and regular spacing of the $\{x_{i}\}$, these can be rewritten as
\begin{eqnarray}
\label{lambdaCn}
\lambda^{C}_{n} & = & - \frac{1}{2V} \sum_{l=1}^{N} e^{ik_{n}x_{l}} \Tintx \Lo(x-x_l) \Eo(x), \\
\label{lambdaDn}
\lambda^{D}_{n} & = & \m{f} \sum_{l=1}^{N} e^{ik_{n}x_{l}} \Tintx \Lo(x-x_l) \Lo(x).
\end{eqnarray}
Let $\Lambda^{C}$ and $\Lambda^{D}$ be the diagonal matrices with entries $\lambda^{C}_{n}$ and $\lambda^{D}_{n}$, and let $R$ be the unitary matrix defined above, with entries 
\begin{equation}
\label{ }
R_{jl} = u^{(l)}_{j} = e^{i k_{l}x_{j}} . \notag
\end{equation}
Then $C=R\Lambda^{C}R^{*}$ and $D=R\Lambda^{D}R^{*}$. Transposition takes eigenvalues to their complex conjugates, so $C^{T}=R\overline{\Lambda^{C}}R^{*}$. From $RR^{*}=I$ and Taylor expansion it follows that $e^{sC}=Re^{s\Lambda^{C}}R^{*}$ and $e^{sC^{T}}=Re^{s\overline{\Lambda^{C}}}R^{*}$. Substitution into Eq. \eqref{lyapCD} then yields a diagonalization of $\var w$:
\begin{eqnarray*}
\label{varwLambda}
\var w & = & R \bigl[ \int_{0}^{\infty} ds \, e^{-s\Lambda^{C}} \Lambda^{D} e^{-s\overline{\Lambda^{C}}} \bigr]  R^{*}  \\
& = & R \Lambda^{w} R^{*},
\end{eqnarray*}
where $\Lambda^{w}$ is the diagonal matrix with entries
\begin{eqnarray}
\label{lambdawn}
\lambda^{w}_{n} & = & \int_{0}^{\infty} ds \, e^{-s\lambda^{C}_{n}} \lambda^{D}_{n} e^{-s\overline{\lambda^{C}_{n}}} \notag  \\
& = & \frac{\lambda^{D}_{n}}{2 Re \lambda^{C}_{n}},
\end{eqnarray}
provided $Re \lambda^{C}_{n} > 0$. Since $D$ is symmetric positive definite (it is, by construction, a physical covariance matrix), we have $\lambda^{D}_{n}$ real and positive for all $n$. Recall that in order for the solution of the Lyapunov equation Eq. \eqref{lyapCD} to be positive definite, all eigenvalues of $C$ must have positive real part, i.e. $Re \lambda^{C}_{n} > 0$ for all $n$. If this physicality condition is satisfied, then the eigenvalues of $\var w$ given by Eq. \eqref{lambdawn} are real and positive. These eigenvalues, with $\lambda^{C}_{n}$ and $\lambda^{D}_{n}$ given by Eq. \eqref{lambdaCn} and \eqref{lambdaDn}, are the variances associated with the independent components of the equilibrium weight distribution. The corresponding eigenvectors are the $u^{(n)}$, with components $u^{(n)}_{j}= e^{ik_{n}x_{j}}$. 

Since $V>0$, the condition for physicality of the covariance is
\begin{eqnarray*}
Re \, \sum_{l=1}^{N} e^{ik_{n}x_{l}} \Tintx \Lo(x-x_l) \Eo(x) & < & 0 \tforall n.
\end{eqnarray*}
This coincides with the condition derived in \cite{williams03} for stability of the \emph{mean} weight state. Roughly speaking, it follows that if there exists an equilibrium weight distribution $P(w)$ (with finite covariance matrix), then the mean of the distribution must be stable. We do not address stability of the equilibrium distribution (or equivalently, stability of all moments of the equilibrium distribution) in the present paper, but a natural conjecture would be that if the equilibrium distribution $P(w)$ exists, then it is necessarily stable.  

From $\var w = R \Lambda^{w} R^{*}$ we can now write down explicit expressions for the equilibrium covariance of any pair of weights:
\begin{eqnarray}
\label{varwjlhom}
\var(w_{j},w_{l})  & = & \sum_{n,m=1}^{N} R_{jn} \Lambda^{w}_{nm} R^{*}_{ml} \notag \\
 & = & \sum_{n=1}^{N} R_{jn}\overline{R_{n}} \lambda^{w}_{n} \notag  \\
 & = & \sum_{n=1}^{N} e^{ik_{n}(x_{j}-x_{l})} \lambda^{w}_{n},
\end{eqnarray} 
with $\lambda^{w}_{n}$ given by Eq. \eqref{lambdawn} and $\lambda^{D}_{n}$, $\lambda^{D}_{n}$ given by Eq. \eqref{lambdaCn} and \eqref{lambdaDn}.

Note that $\var(w_{j},w_{l})$ depends on $j$ and $l$ only via the difference $(x_{j}-x_{l}) \text{\,mod\,} T$, due to periodicity and translational invariance of the architecture for homogeneous parameters. Also, the covariance of the weights depends only on the associative part $\L$ of the learning rule, since the nonassociative part $\alpha$ does not appear in Eq. \eqref{varwjlhom}. This is not surprising, since the role of $\alpha$ is essentially analagous to that of a constant externally applied force in a physical system. Such a force changes the position of the equilibrium, but does not alter the dynamics around the equilibrium.

\subsection{Confinement}
Our derivation of the moment hierarchy relations Eq. \eqref{recurrmuvec} relied on the assumption that the equilibrium weight distribution was negligible on the ``tails'' of the piecewise linear postsynaptic gain function $f$. This places a constraint on the mean $\m{U}(x)$ and diagonal variance $\var(U(x),U(x))$ of the postsynaptic  potential: they must be such that the mean is a large number of standard deviations away from the tails. For each $x$, let $r(x)$ be the standard deviation of $U(x)$ divided by the distance from $\m{U}(x)$ to the nearest tail, i.e. to $V-\theta$ or $-V-\theta$. 
The parameter $r(x)$ will be referred to as the \emph{confinement parameter} for the system. The confinement condition holds provided $\m{U}(x)$ is in the interval $(-V-\theta,V-\theta)$ and $r(x) \ll 1$, for all $x$.  

We now argue that by adjusting only the rates of nonassociative and associative learning, the confinement condition can always be satisfied. Multiplying the associative learning rule by a positive scalar factor $\beta$ and both nonassociative and associative components by a positive scalar factor $\lambda$, we have weight changes given by
\begin{equation}
\label{deltawvecbetalambda}
\tri w(t)=\begin{cases} \lambda\alpha+\beta\Lo(x), \text{ density }
\Tinv\tif(x,w(t)) \\ \lambda\alpha, \text{ probability }  1-\Tintx\tif(x,w(t)).
\end{cases}    
\end{equation}
The ratio of associative to nonassociative learning rate is parametrized by $\beta$, while the overall learning rate is parameterized by $\lambda$. Now it was shown in \cite{williams03} that in the case of homogeneous parameters, under certain mild conditions, the equilibrium mean weight vector has the property that $\m{U}(x)$ is approximately constant (i.e. the equilibrium is an approximate negative image state). Hence $\m{f}$ in Eq. \eqref{meanfmeanw} is approximately constant. If it were exactly constant then Eq. \eqref{meanfmeanw} (for homogeneous parameters) would yield, after cancelling $\lambda$ on top and bottom,
\begin{eqnarray*}
\m{f} =  \frac{-\alpha}{\alpha+\frac{\beta}{T}\int dx \, \Lo(x)}.
\end{eqnarray*}
Provided $\alpha$ and $\int dx\, \Lo(x)$ have opposite sign (shown in \cite{williams03} to be necessary for existence of a negative image equilibrium) the right hand side of this equation can be made to have any desired value by appropriate choice of $\beta>0$. Hence $\m{f}$ can be made to have any desired value by appropriate choice of $\beta$; in particular, a range of $\beta$ exists for which $\m{f}$ falls in the open interval $(f(-V-\theta),f(V-\theta))$. Since $f$ is invertible for arguments in $(-V-\theta,V-\theta)$ and $\m{f}=f(\m{U})$, it follows that by appropriate choice of $\beta$, $\m{U}$ can be made to have any value in $(-V-\theta,V-\theta)$. Since $\m{f}(x)$ approximately constant implies $\m{U}$ approximately constant, it follows that the mean postsynaptic potential $\m{U}(x)$ can always be made to lie between the tails, for all $x$. 

It remains to show that the diagonal variance $\var(U(x),U(x))$ can be made sufficiently small so that the distribution of $U(x)$ is negligible on the tails. We do this by holding $\beta$ fixed and varying $\lambda$. Since the matrix $C$ is proportional to $\lambda$ and the matrix $\var \tri w$ is proportional to $\lambda^2$, it follows from Eq. \eqref{lyapvarwalmost} that $\var w$, and hence $\var U$ from Eq. \eqref{}, is proportional to $\lambda$. In particular, $\var(U(x),U(x))$ can be made arbitrarily small by taking $\lambda$ sufficiently small. 

Thus, by appropriate choice of $\beta$ and $\lambda$, the confinement condition can always be satisfied. The value of $\beta$ determines the location of the mean postsynaptic potential, and the value of $\lambda$ determines the width of the distribution around the mean. The latter fact, that the width of the equilibrium distribution of the postsynaptic potential is proportional to the overall learning rate, has direct behavioral relevance to the mormyrid fish, since it implies a tradeoff between speed of adaptation and accuracy of the adapted state\footnote{The fact that the variance is proportional to the learning rate is also true for inhomogeneous parameters, by the same argument. But the confinement of the mean postsynaptic potential $\m{U}(x)$ is unclear in that case, because the equilibrium is not necessarily an approximate negative image. Further work is required to characterize the equilibrium for inhomogeneous parameters.}.

\subsection{Dense Spacing Limit}
 
In the architecture of mormyrid ELL, the spacing $\delta$ between presynaptic spike times is much less than the widths $\tau_{E}$, $\tau_{L}$ of the PSP $\E$ and learning rule $\L$.
In the dense spacing limit the set of discrete weights per unit time $\{w_{i}/T\}$ corresponding to presynaptic spikes at times $\{x_{i}\}$ becomes a continuum weight density $\W(y)$, with weight $\W(y) dy$ corresponding to presynaptic spike times between $y$ and $y+dy$. Sums over $x_{i}$ are replaced by integrals over $y$.
The matrices $C$ and $D$ in Eq. \eqref{lyapCD} become infinite dimensional, with eigenvalues $\lambda_{n}^{C}$, $\lambda_{n}^{D}$ given by
\begin{eqnarray}
\label{lambdaCndensealmost}
\lambda^{C}_{n} & = & - \frac{1}{2VT} \int_{0}^{T} dy\, e^{ik_{n}y} \int_{0}^{T} dx\,  \Lo(x-y) \Eo(x), \\
\label{lambdaDndensealmost}
\lambda^{D}_{n} & = & \frac{\m{f}}{T}  \int_{0}^{T} dy\, e^{ik_{n}y} \int_{0}^{T} dx\, \Lo(x-y) \Lo(x).
\end{eqnarray}
for $n=0,1,\ldots$.
We introduce some useful notation. Let $\FT{h}$ be the sequence of Fourier coefficients for a function $h$ on $[0,T]$, given by $\FT{h}_{n} =  \int_{0}^{T}dy\, e^{ik_{n}y} h(y)$ with $k_{n}=2\pi n/T$, $n=0,1,\ldots$. Let $*_{T}$ denote convolution on the interval $[0,T]$, $(g *_{T} h)(x) = \int_{0}^{T} dy\,  g(x-y) h(y)$. Let $\widetilde{h}$ denote the horizontal reflection of $h$, $\widetilde{h}(y)=h(-y)$. Then Eq. \eqref{lambdaCndensealmost} and \eqref{lambdaDndensealmost} can be written as 
\begin{eqnarray*}
\label{lambdaCndensenotquite}
\lambda^{C}_{n} & = & - \frac{1}{2VT} \FT{\Lo *_{T} \widetilde{\Eo}}_{n}, \\
\label{lambdaDndensenotquite}
\lambda^{D}_{n} & = & \frac{\m{f}}{T} \FT{\Lo *_{T} \widetilde{\Lo}}_{n}.
\end{eqnarray*}
Now we invoke the Fourier convolution theorem $\FT{g*h}=\FT{g}\FT{h}$, and the fact that $\FT{\widetilde{g}}=\overline{\FT{g}}$, where $\overline{z}$ denotes the complex conjugate of $z$. This gives
\begin{eqnarray}
\label{lambdaCndense}
\lambda^{C}_{n} & = & - \frac{1}{2VT} \FT{\Lo}_{n} \overline{\FT{\Eo}_{n}},  \\
\label{lambdaDndense}
\lambda^{D}_{n} & = & \frac{\m{f}}{T} \FT{\Lo}_{n} \overline{\FT{\Lo}_{n}}.
\end{eqnarray}
The eigenvalues of the weight covariance are therefore
\begin{eqnarray}
\label{lambdaWndense}
\lambda^{\W}_{n} & = & \frac{\lambda^{D}_{n}}{2 Re \, \lambda^{C}_{n}}  \notag \\
 & = & - \m{f}V \frac{\FT{\Lo}_{n}\overline{\FT{\Lo}_{n}}}{Re\,\bigl[ \FT{\Lo}_{n} \overline{\FT{\Eo}_{n}}\bigr] }.
\end{eqnarray}
It follows that the covariance of $\W(y)$ and $\W(z)$ is
\begin{eqnarray}
\label{varWyzdense}
 \var (\W(y),\W(z)) = \sum_{n=0}^{\infty} e^{ik_{n}(y-z)} \lambda^{W}_{n}   \notag   \\
 =  - 2 \pi \m{f}V \FTinv{\frac{\FT{\Lo}\overline{\FT{\Lo}}}{Re\, \bigl[ \FT{\Lo} \overline{\FT{\Eo}} \bigr]} } (y-z),
\end{eqnarray}
where $\FTinv{h}(x)=(1/2 \pi)\sum_{n=0}^{\infty} e^{ik_{n}x} h_{n}$ is the inverse Fourier transform on $[0,T]$. The covariance of the postsynaptic potential is then
\begin{eqnarray}
 \var U(y,z) &=&  \int_{0}^{T}dx \int_{0}^{T}dx'\,  \Eo(y-x) \var \W(x,x') \Eo(z-x') \notag \\
 & =&  - 2 \pi \m{f}V   \int_{0}^{T}dx \int_{0}^{T}dx'\,  \Eo(y-x)  \Eo(z-x') \notag \\
 && \times \FTinv{\frac{\FT{\Lo}\overline{\FT{\Lo}}}{Re\, \bigl[ \FT{\Lo} \overline{\FT{\Eo}} \bigr]} } (x-x').
\end{eqnarray}
One special case is worth noting: suppose the PSP and learning rule have identical functional form, i.e. are proportional to one another, $\L(x)=c\E(x)$ for some (real) constant $c$. Then we have
\begin{equation}
\label{ }
\FTinv{\frac{\FT{\Lo}\overline{\FT{\Lo}}}{Re\, \bigl[ \FT{\Lo} \overline{\FT{\Eo}} \bigr] } } (x) = \FTinv{c}(x) = \frac{c}{2 \pi} \delta(x).  \notag 
\end{equation}
where $\delta(x)$ is the Dirac delta function. For such a learning rule the covariance of the weight density is
\begin{equation}
\label{varWyzLpropE}
\var (\W(y),\W(z)) = -\m{f} V c \, \delta(y-z).
\end{equation}
In particular, the covariance (and hence the correlation) of $\W(y)$ and $\W(z)$ is zero for $y \ne z$, hence weights corresponding to different presynaptic spike times are statistically independent. This is surprising, since the coupling of weights through the PSP $\E$ and learning rule $\L$ has some nonzero ``range'', given roughly by the widths of $\E$ and $\L$, and within this range one would expect the weights to necessarily have some nonzero correlation. The result just derived says that in certain exceptional cases this correlation may  vanish. The result was derived in the dense spacing limit, but can be expected to hold approximately for the physical case of discrete spacing, and also to hold approximately for $\L$ not quite proportional to $\E$; this will be verified in the examples calculated below. Given that the best current experimental measurement of the learning rule in mormyrid ELL \cite{Bell97a} is not inconsistent with $\E$ and $\L$ having the same functional form, this vanishing correlation phenomenon may have biological relevance.

\section{Examples}\label{sec.examples}

\begin{figure}
\begin{center}
\includegraphics[width=\figwidth]{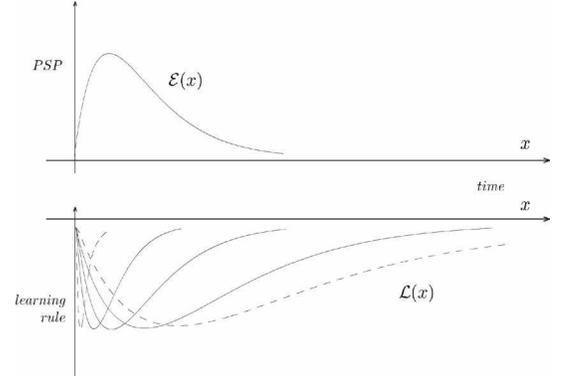}
\caption{PSP and learning rules used in the examples. Stability requires $3-2\sqrt{2}<
  \tau_L/\tau_E<3+2\sqrt{2}$. Stable examples are drawn with solid
  lines; endpoints of the stable interval are drawn with dashed
  lines. Arbitrary units.}
\label{figalphaLE}
\end{center}
\end{figure}

We now compute the equilibrium weight covariances for a class of PSPs and learning rules consistent with those measured in mormyrid ELL, assuming homogeneous parameters. The PSP we take to be an excitatory alpha function of width $\tau_{E}$, and the learning rule we take to be alpha function, depressive, and pre-before-post, of width $\tau_{L}$:
\begin{eqnarray}
\E(x) & = & \tau_{E}^{2}e^{-x/\tau_{E}} H(x), \label{Eee}\\
\L(x) & = & -\tau_{L}^{2}e^{-x/\tau_{L}} H(x),  \label{Lee}
\end{eqnarray} 
where $H(x)$ is the Heaviside function, $H(x)=1$ if $x \ge 0$ and $0$ otherwise (Fig \ref{figalphaLE}). In the above expressions both $\E$ and $\L$ have been normalized to unit area, but to ensure confinement of the postsynaptic potential, the learning rule $\L$ (and hence the size of the learning steps) must be made sufficiently small so that the confinement condition is satisfied. 

It was shown in \cite{williams03} that in order for the mean weight dynamics to be stable near the (negative image) equilibrium, the time constants  $\tau_{E}$ and $\tau_{L}$ must satisfy
\begin{eqnarray*}
3-2\sqrt{2} < \frac{\tau_{L}}{\tau_{E}} < 3+2\sqrt{2}.
\end{eqnarray*}

\begin{figure}
\begin{center}
\newcommand{\sfw}{\figwidth}
\subfigure[]{\includegraphics[width=\sfw]{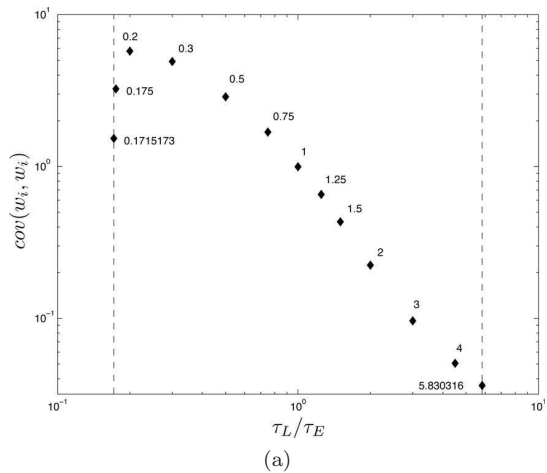}}
\caption{Diagonal variance of weights, for alpha function $\E$ and $\L$ and for various values of $\tau_L/\tau_E$. The larger of $\tau_L$ and $\tau_E$ was taken to be $0.2T$ in all cases. Diagonal variance vs $\tau_L/\tau_E$, log-log plot. Dotted lines indicate the boundary of the stable interval, $\tau_L/\tau_E=3\pm 2\sqrt{2}$. Dimensionless units.}
\label{figdiagcovw}
\end{center}
\end{figure}

\begin{figure}
\begin{center}
\newcommand{\sfw}{\figwidth}
\subfigure[]{\includegraphics[width=\sfw]{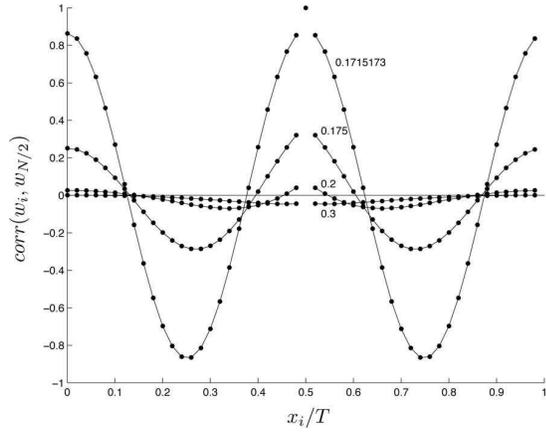}}
\subfigure[]{\includegraphics[width=\sfw]{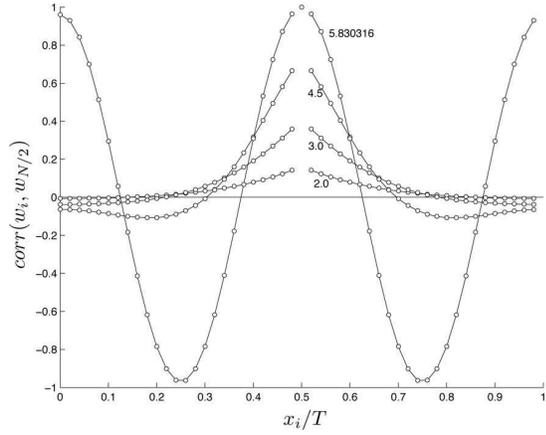}}
\subfigure[]{\includegraphics[width=\sfw]{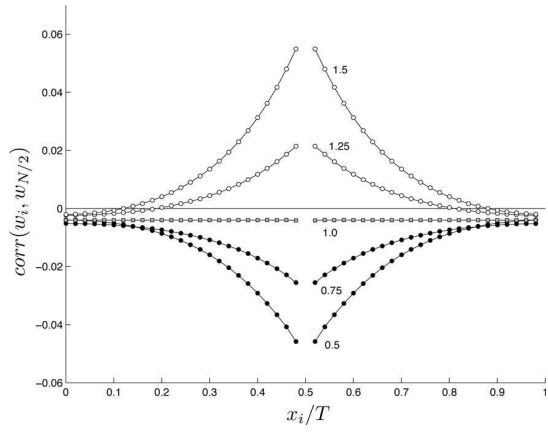}}
\caption{Correlation of weights, for alpha function $\E$ and $\L$ and for various values of $\tau_L/\tau_E$. The larger of $\tau_L$ and $\tau_E$ was taken to be $0.2T$ in all cases. Curves are labelled by the value of $\tau_L/\tau_E$, and for clarity curves are not joined to the point (0.5,1) which all curves have in common. (a) Correlation of $w_i$ with $w_{N/2}$, versus $x_i/T$, for $\tau_L/\tau_E$ significantly less than $1$. (b) Same, for $\tau_L/\tau_E$ significantly greater than $1$. (c) Same, for $\tau_L/\tau_E$ near $1$, with expanded vertical scale. Dimensionless units.}
\label{figcovw}
\end{center}
\end{figure}

For $\tau_L/\tau_E$ in this stable range, we calculated the equilibrium covariance of the synaptic weights and of the postsynaptic potential, and verified our predictions by direct Monte Carlo simulation of the underlying random walk. The number of presynaptic cells was taken to be $N=50$, and to ensure that the confinement condition was well satisfied, the rates of nonassociative and associative learning were adjusted so that the confinement parameter was $r(x)=0.2$ for all $x$ (i.e. the tails were 5 standard deviations away from the mean postsynaptic potential). By translational symmetry for homogeneous parameters, the diagonal variances $\var(w_i,w_i)$ are independent of $i$, and the off-diagonal covariance $\var(w_i,w_j)$ depends only on $(x_i-x_j)\text{\,mod\,} T$. The covariance matrix is then  completely described by the diagonal variance (a single number) and the correlation of weight $w_i$ with the ``midpoint'' weight $w_{N/2}$, for $i=1,2,\ldots,N$; the correlation in this case is just the covariance normalized by the diagonal variance. The diagonal variance is shown in Fig. \ref{figdiagcovw}, and the correlation is shown in Fig. \ref{figcovw}, for various values of $\tau_{L}/\tau_{E}$ between $3-2\sqrt{2}$ and $3+2\sqrt{2}$. Note the approximate vanishing of off-diagonal correlation for $\tau_L/\tau_E$ near $1$, as expected from the analytic calculation in the dense-spacing limit. The manner in which the correlation deviates from an approximate delta function as $\tau_L/\tau_E$ deviates from $1$ also shows an interesting pattern: for $\tau_L/\tau_E$ slightly greater than $1$, the near-diagonal (near-neighbor) correlation is positive, while for $\tau_L/\tau_E$ slightly less than $1$, the near-neighbor correlation is negative. But for $\tau_L/\tau_E$ substantially greater than or less than $1$, the near-neighbor correlation is positive in both cases. The magnitude of off-diagonal correlation tends to increase as $\tau_L/\tau_E$ moves away from $1$ in either direction. Near the limits of the stable range of $\tau_L/\tau_E$, the near-neighbor correlation is close to $1$ and the ``antipodal'' correlation (correlation with weights a half period away) is close to $-1$. Such strong long-range correlation/anticorrelation was also observed numerically in \cite{Roberts00b} in mean weight dynamics for parameters near the boundary of the stable region, with breakdown of stability being characterized by the appearance of travelling waves.

\begin{figure}
\begin{center}
\newcommand{\sfw}{\figwidth}
\subfigure[]{\includegraphics[width=\sfw]{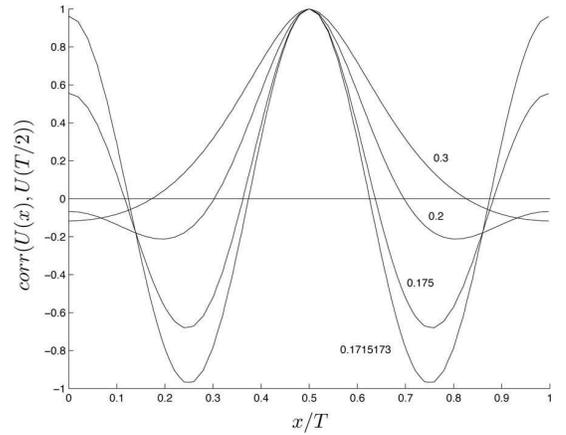}}
\subfigure[]{\includegraphics[width=\sfw]{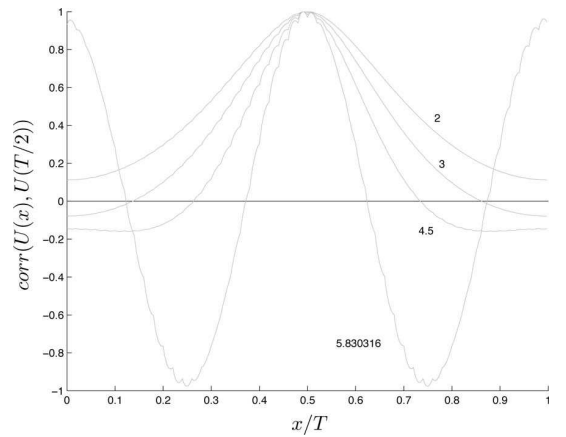}}
\subfigure[]{\includegraphics[width=\sfw]{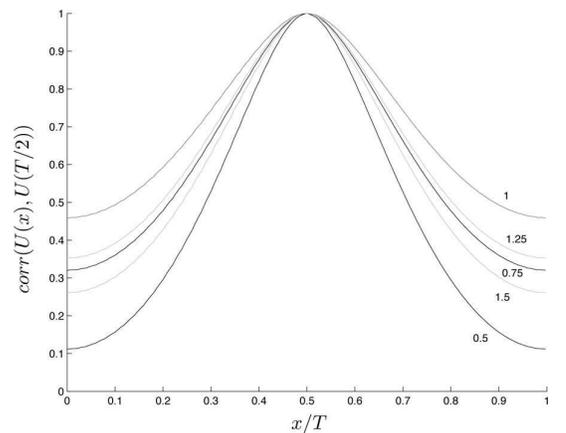}}
\caption{Correlation of postsynaptic potential, for alpha function $\E$ and $\L$ and for various values of $\tau_L/\tau_E$. The larger of $\tau_L$ and $\tau_E$ was taken to be $0.2T$ in all cases. (a) Correlation of $U(x)$ with $U(T/2)$, versus $x/T$, for $\tau_L/\tau_E$ significantly less than $1$. (b) Same, for $\tau_L/\tau_E$ significantly greater than $1$. (c) Same, for $\tau_L/\tau_E$ near $1$, with expanded vertical scale. Curves are labelled by the value of $\tau_L/\tau_E$. Dimensionless units. }
\label{figcovU}
\end{center}
\end{figure}

The correlation of the postsynaptic potential is shown in Fig. \ref{figcovU}. For $\tau_L/\tau_E$ near $1$ the correlation is everywhere positive. As $\tau_L/\tau_E$ deviates from $1$, the correlation decreases, and long-range anti-correlations appear. As $\tau_L/\tau_E$ deviates still further, the anti-correlation decreases in range and increases in magnitude, and a positive long-range correlation appears. For $\tau_L/\tau_E$ near the limits of the stable range, the mid-range and long-range (antipodal) correlations approach $-1$ and $+1$, respectively, similar to the behavior of the synaptic weight correlation. The ``scalloped'' appearance of these curves for large $\tau_L/\tau_E$ is due to $\tau_E$ being not much larger than the spacing $\delta=T/50$ between presynaptic spike times, resulting in only marginal overlap of adjacent PSPs. For fixed PSP width $\tau_E$, such scalloping should vanish as the spacing of presynaptic spike times goes to zero. It is believed [C.C. Bell, private communication] that in mormyrid ELL the spacing of presynaptic spike times is sufficiently dense that this scalloping would be insignificant.

\begin{figure}
\begin{center}
\newcommand{\sfw}{\figwidth}
\subfigure[]{\includegraphics[width=\sfw]{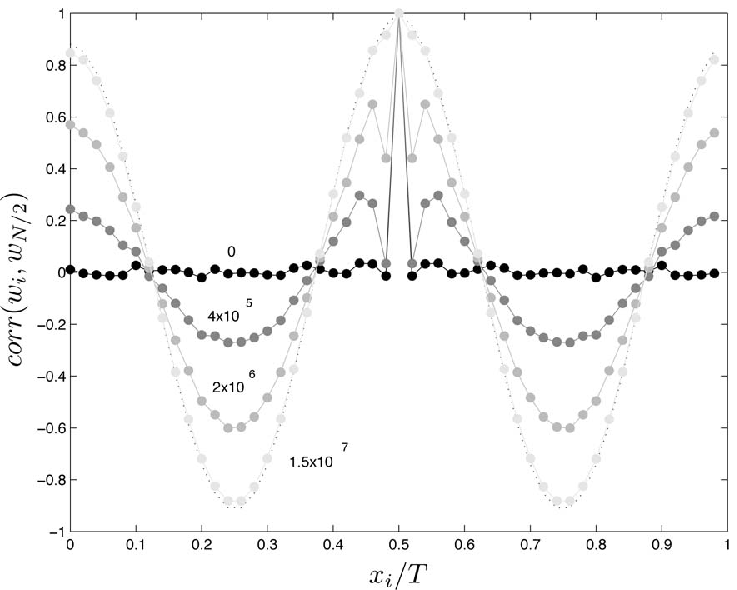}}
\subfigure[]{\includegraphics[width=\sfw]{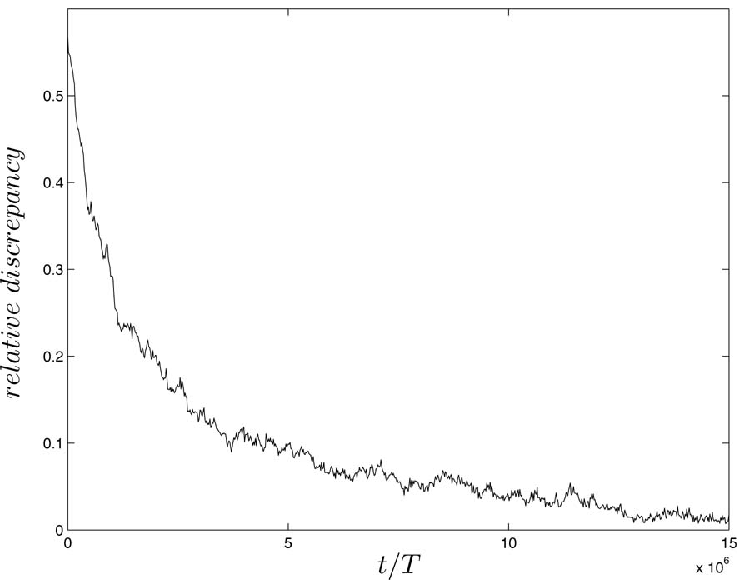}}
\caption{Convergence of weight correlation to predicted equilibrium values in Monte Carlo simulation, for $L/E=5.81$, $N=50$, confinement parameter $=0.2$. (a) Time-evolution of population-averaged correlation; curves labelled by time, $t/T$. Dotted curve indicates prediction. (b) Relative discrepancy between predicted and actual correlation, vs. time $t/T$. Dimensionless units.}
\label{figsimulation}
\end{center}
\end{figure}

Comparison with direct Monte Carlo simulation of the random walk revealed excellent agreement with prediction, provided confinement was well satisfied; results for $\tau_{L}/\tau_{E}=5.814$, near the upper end of the stable range, are shown in Fig. \ref{figsimulation}. As above, nonassociative and associative learning rates were adjusted so that the confinement parameter $r(x)$ was $0.2$ for all $x$ (i.e. the tails were five standard deviations away from the equilibrium mean). Weights were taken to be initially uncorrelated, with mean equal to the predicted mean and variance equal to the predicted (diagonal) variance; the initial correlation was then the discrete Dirac delta function. To quantify convergence we used the mean absolute value of the relative discrepancy between the predicted and actual (ensemble mean) correlation. Translation invariance of the correlation allowed us to reduce the size of fluctuations in the simulation estimate by averaging not just over the ensemble but also over the population of $N=50$ weights in each member of the ensemble\footnote{Although the predicted correlation is translation invariant, the fluctuations around the prediction are not necessarily uncorrelated. For our purposes this is harmless; it simply means that we don't obtain as large a reduction in fluctuation size by population averaging as we would by using a $50$-times larger ensemble.}. Using this measure, the correlation in the simulation converged to within $1$ to $2$ percent of the predicted correlation in approximately $10^7$ timesteps (Fig. \ref{figsimulation}).

\section{Summary}\label{sec.summary}

Since changes in synaptic weights in STDP are due to temporally discrete events (spikes or spike pairs), the dynamics of such plasticity, in the presence of noise, is naturally modelled as a discrete-time random walk. There is a large body of mathematical technique for the analysis of such processes \cite{hughes}.

From the weight dynamics expressed as a random walk one can write down a master equation for the time evolution of the weight probability distribution. From the master equation we obtain a functional equation for the equilibrium weight distribution. Taking the Fourier transform of this equation yields a differential equation for the characteristic function of the equilibrium distribution, and Taylor expansion then yields a hierarchy of recurrence relations for the equilibrium moments. From the moments of the equilibrium weight distribution we also obtain the moments of the postsynaptic membrane potential. 

For the case of a single weight, we explicitly calculate moments up to fourth order. The distribution is shown to be generically non-Gaussian, but the skew and kurtosis approach Gaussian values as the learning rate (size of steps) goes to zero. 
 
For the case of multiple weights we explicitly calculate moments up to second order. The mean weight vector satisfies a simple matrix-vector equation, which is equivalent to the condition that the mean step in the equilibrium state is zero, for all weights. The weight covariance matrix satisfies a Lyapunov equation. An explicit solution to this equation, in the form of a matrix integral, is obtained. For this solution to be the covariance matrix of some probability distribution it must be positive definite, which imposes a constraint on the PSP $\E$ and the associative learning rule $\L$. 

For the case of multiple weights with homogeneous parameters, further analytical progress can be made. The Lyapunov equation for the weight covariance matrix can be fully diagonalized, and the covariance of any pair of weights found in closed form. From this we also obtain explicit expressions for the covariance of the postsynaptic potential between any pair of times. The physicality condition, that the weight covariance matrix be positive definite, takes an especially simple form in this case, closely related to the condition derived in \cite{williams03} for stability of the mean weight state. 

In the limit of dense spacing of presynaptic spike times, the expression for the weight covariance is further simplified. In the special case where $\E$ and $\L$ have the same functional form, we find, surprisingly, that weights corresponding to distinct presynaptic spike times are statistically independent. This result can be expected to hold approximately for discrete presynaptic spike times, and for learning rules not quite identical to $\E$ in functional form. 

Numerical calculation of the equilibrium weight covariance and postsynaptic potential covariance was carried out for a class of examples relevant to mormyrid ELL: both $\E$ and $\L$ alpha function in form, with $\E$ excitatory and $\L$ depressive pre-before-post. For the synaptic weights, off-diagonal correlation is near zero for $\tau_L/\tau_E=1$, and tends to increase in magnitude as $\tau_L/\tau_E$ moves away from $1$. Values of $\tau_L/\tau_E$ near the boundary of the stable range show large long-range anticorrelations. The correlation of the postsynaptic potential is everywhere positive for $\tau_L/\tau_E=1$, but long-range anticorrelations develop as $\tau_L/\tau_E$ moves away from $1$. These numerical predictions were found to be in excellent agreement with direct Monte Carlo simulation of the underlying random walk.

\begin{acknowledgments}
We would like to thank Dr. Gerhard Magnus, Dr. Nathaniel Sawtell, and
the members of Dr. Curtis Bell's lab for insightful discussions.
This material is based upon work supported by the National Science
Foundation under Grant No. IBN-0114558, and by the National Institute 
of Mental Health under Grant No. R01-MH60364.
\end{acknowledgments}


\end{document}